	\documentclass[twoside,journey]{IEEEtran}
	\usepackage{makecell}
	\usepackage{array}
	\usepackage{graphicx,amssymb,amsmath}
	\usepackage{multicol}
	\usepackage[noadjust]{cite}
	\usepackage{setspace}
	\usepackage{subfigure}
	\usepackage{graphicx}
	\usepackage{float}
	\usepackage {url}
	\usepackage{stfloats}
	\usepackage{amsthm,pifont}
	\usepackage{flushend}
	\usepackage{cases,subeqnarray}
	\usepackage{bm,multirow,bigstrut}
	\usepackage{amsmath, amsthm, amssymb}
	\usepackage{textcomp}
	\usepackage{latexsym,bm}
	\usepackage{booktabs}
	\usepackage{xcolor}
	\usepackage{mathtools}
	\usepackage{dsfont}
	\usepackage{extarrows}
	\usepackage{epsfig}
	\usepackage{epsfig}
	\usepackage{epstopdf}
	\usepackage[noend]{algpseudocode}
	\usepackage{algorithmicx,algorithm}
	\theoremstyle{plain}

	\theoremstyle{plain}
	\newtheorem{rem}{Remark}

	\newtheorem{prop}{Proposition}
	\usepackage{amsmath}
	\newtheorem{theorem}{Theorem}
	\newtheorem{lemma}{Lemma}
	\IEEEoverridecommandlockouts
\begin{document}
	\title{Performance Analysis and Optimization for Jammer-Aided Multi-Antenna UAV Covert Communication}
	\author{Hongyang Du, Dusit~Niyato,~\IEEEmembership{Fellow,~IEEE}, Yuan-ai~Xie, Yanyu~Cheng, Jiawen~Kang$^*$, and Dong~In~Kim,~\IEEEmembership{Fellow,~IEEE}
	
	\thanks{H.~Du is with the School of Computer Science and Engineering, the Energy Research Institute @ NTU, Interdisciplinary Graduate Program, Nanyang Technological University, Singapore (e-mail: hongyang001@e.ntu.edu.sg).}
	\thanks{D. Niyato and J. Kang are with the School of Computer Science and Engineering, Nanyang Technological University, Singapore (e-mail: dniyato@ntu.edu.sg; kavinkang@ntu.edu.sg)}
	\thanks{Yuan-ai Xie is with the Institute of Electrical Engineering, Yanshan University, Qinhuangdao 066004, China. (emails: xieyuan\_ai@163.com)}
	\thanks{Y. Cheng is with Alibaba-NTU Singapore Joint Research Institute, Nanyang Technological University, Singapore 639798 (e-mail: yanyu.cheng@ntu.edu.sg).}
	\thanks{D. I. Kim is with the Department of Electrical and Computer Engineering, Sungkyunkwan University, Suwon 16419, South Korea (e-mail: dikim@skku.ac.kr)}
	}
	\maketitle
	\vspace{-2cm}
	\begin{abstract}
	Unmanned aerial vehicles (UAVs) have attracted a lot of research attention because of their high mobility and low cost in serving as temporary aerial base stations (BSs) and providing high data rates for next-generation communication networks. {\color{black} To protect user privacy while avoiding detection by a warden, we investigate a jammer-aided UAV covert communication system, which aims to maximize the user's covert rate with optimized transmit and jamming power. The UAV is equipped with multi-antennas to serve multi-users simultaneously and enhance the Quality of Service.} By considering the general composite fading and shadowing channel models, we derive the exact probability density (PDF) and cumulative distribution functions (CDF) of the signal-to-interference-plus-noise ratio (SINR). The obtained PDF and CDF are used to derive the closed-form expressions for detection error probability and covert rate. Furthermore, the covert rate maximization problem is formulated as a Nash bargaining game, and the Nash bargaining solution (NBS) is introduced to investigate the negotiation among users. To solve the NBS, we propose two algorithms, i.e., particle swarm optimization-based and joint two-stage power allocation algorithms, to achieve covertness and high data rates under the warden's optimal detection threshold. {\color{black} All formulated problems are proven to be convex, and the complexity is analyzed. The numerical results are presented to verify the theoretical performance analysis and show the effectiveness and success of achieving the covert communication of our algorithms.}
	\end{abstract}
	\begin{IEEEkeywords}
	Covert communication, multi-antenna UAV, performance analysis, optimization, Bargaining game.
	\end{IEEEkeywords}
	
	%=============================================================
	\section{Introduction}	\label{sec:introduction}
	Wireless communication networks have always required a high data rate and secure transmission. As sixth-generation networks aspire for higher capacity, lower latency, and higher user density, the number of mobile users and devices is rapidly rising, which poses great challenges in the near-future development of Internet-of-things (IoT). In remote areas without terrestrial infrastructures and disaster-affected areas with damaged infrastructures, efficient communication service for ground users and devices cannot be guaranteed. Furthermore, since each IoT device has a restricted transmission range, the signal cannot travel a long distance.
	Due to these limitations, common ground base stations (BSs) may be unable to cover all IoT devices and gather data effectively. As a result, a new paradigm is urgently required to enhance the cellular network's quality of service (QoS). 
	
	{\color{black} Because of the flexible operation and large coverage \cite{zhou2021three}, unmanned aerial vehicle (UAV)-aided communications are regarded as one of the most promising techniques for future networks \cite{li2019secure,lin2021supporting}.} By acting as temporary aerial BSs, UAVs can perform a broad range of activities, including disaster rescue support, real-time surveillance, and data collection \cite{nguyen2021uav,cui2018robust}. To meet the ever-increasing demands, it is potential to adopt multi-antennas at the UAV to improve the channel quality \cite{yang2021performance}. {\color{black} The multi-antenna access and high mobility of the UAV improve the system performance significantly, but several challenges remain. For example, because of the wide coverage and large-scale connection provided by UAVs, open networks are susceptible to eavesdropping and various attacks by malicious adversaries \cite{zhu2020millimeter,jiang2021covert}. Since UAVs are commonly used to transmit private data, it is critical to design a secure UAV-aided system \cite{chen2021multi,jiang2021resource,zhang2020optimized,shahzad2018achieving}.}
	
	Several related works consider enhancing the security of UAV networks \cite{yoon2017security,atoev2019secure,zhang2019securing,cui2018robust}, which can be divided into two major categories, i.e., cryptographic methods \cite{yoon2017security,atoev2019secure} and physical layer security techniques \cite{nguyen2021uav,zhang2019securing,cui2018robust}. {\color{black} However, several practical situations require not only the security of the transmitted content but also the covertness of the transmission behavior, which has not been addressed and is also desirable for UAV networks.} The reason is that once the transmission behavior of a transmitter is detected by malicious users, its location information is exposed, which makes the transmitter vulnerable to physical attacks \cite{zhou2019seeing}. Fortunately, the emerging covert communication technique can effectively hide the existence of a wireless transmission, i.e., avoiding a wireless transmission being detected by the adversary, called warden.
	
	Existing works indicate that covert communication can be achieved through random noises \cite{wang2021energy} or interference \cite{shahzad2018achieving}. To degrade the detection performance of the warden, a friendly jammer is introduced to assist the communication by actively generating jamming signals. Unlike the random noise behaving uncontrollably, the interference can be controlled effectively by optimizing the jammer's power. Nevertheless, subject to the power constraint and the minimum covert-rate requirement of users, the friendly jammer aims to use less power to help each user achieve a higher covertness. The contradicted and independent individual objectives of user-jamming antenna pairs impel us to study how users bargain and negotiate with each other to achieve their covert communication. Hence, it is natural to apply game theory to balance the objectives among different pairs \cite{osborne1994course,han2012game,zheng2016resource}. Accordingly, to design the transmit and jamming power allocation and achieve the optimal secure UAV-aided communication, a Nash bargaining approach is proposed to obtain the Nash bargaining solution (NBS) \cite{nash1950the}. Thus, an agreement can be achieved by players, i.e., $K$ users, efficiently and fairly, given the NBS's five axioms of pareto optimality (PAR), individual rationality (IR), independent of expected utility representations (INV), independence of irrelevant alternatives (IIA) and symmetry (SYM) \cite{nash1950the}.
	%Because our problem and solution are equivalent to standard Nash bargaining problem, the five axioms are achievable intrinsically.
	
	In this paper, we adopt two more practical and accurate channel models, namely Fisher–Snedecor $\mathcal{F}$ \cite{yoo2017the} and Fluctuating Two-Ray (FTR) models \cite{zhang2017new}, than traditional Nakagami-$m$ \cite{cheng2020downlink} and Rician models \cite{liu2021channel} to conduct a detailed performance analysis about the jammer-aided multi-antenna UAV covert communication system. Based on the obtained analytical results, bargaining game theory is used to determine the optimal transmit and jamming power allocation scheme. The main contributions are summarized as follows:
	{\color{black}\begin{itemize}
	\item We investigate a jammer-aided multi-antenna UAV covert communication system. By modeling the ground channel links as Fisher-Snedecor $\mathcal{F}$ fading and the air-to-ground channel links as FTR fading, the exact probability density function (PDF) and cumulative distribution function (CDF) of the end-to-end signal-to-interference-plus-noise ratio (SINR) are derived. To the best of our knowledge, this is the first study that analyzes a jammer-aided UAV covert communication system over composite fading and shadowing with accurate and tractable generalized channel models.
	
	\item The closed-form expressions for the detection error probability and covert rate are derived. To maximize the user's covert rate with limited transmit and jamming power, the optimization problem is formulated as a Nash bargaining game (NBG) with the help of the derived performance metrics. Furthermore, the NBS is introduced to investigate the transmit and jamming power resource negotiation among the players, i.e., users in multi-antenna UAV-aided covert communication networks, to reach an agreement that is efficient and fair. Moreover, with the help of characteristics of generalized hypergeometric functions, we investigate the optimization problem's convergence based on analytical expressions. The convergences of all problems formulated in this paper are proven, and the complexity is formally analyzed.
	
	\item We propose two algorithms, i.e., particle swarm optimization (PSO)-based power allocation (PPA) and joint two-stage power allocation (JTPA) algorithms, to solve the formulated problem. To demonstrate the robustness of the proposed algorithms, we assume that the warden can find the optimal detection threshold, i.e., the warden can do the most harm to the covertness. A detection error probability minimization algorithm for the warden is also presented. In the proof of convergence, we study an integral equation that has not been analyzed in any function books, e.g., \cite{Prudnikov1986Integrals,gradshteyn2007,web}. The obtained results and approximations are useful in the performance analysis of wireless communications systems.
	\end{itemize}}
	The remainder of the paper is organized as follows: In Section \ref{sec:FL}, we introduce the jammer-aided multi-antenna UAV covert communication, and derive exact statistics for the end-to-end SINR. In Section \ref{S3}, closed-form performance metrics, such as detection error probability and covert rate, are derived. With help of the derived metrics and bargaining game theory, we formulate the optimization problem as a NBG to maximize the user's covert rate in Section \ref{section4}. Section \ref{S5} presents the PPA and JTPA algorithms.
	Numerical results and Monte-Carlo simulations are presented in Section \ref{S6} to verify the accuracy of our analysis and proposed power allocation algorithms. Finally, Section \ref{S7} concludes this paper.
	
	\begin{table}[htbp]\label{table1}
	\caption{Mathematical Notations and Functions}
	\centering
	{\small\begin{tabular}{l|p{6cm}}
	\toprule
	\hline
	$\textrm{Pr}\{\cdot\}$ & Probability function \\
	$\bm{a}^T$ &   Transpose of vector $\bm{a}$ \\
	$K$ & The average power ratio of the dominant wave to the scattering multipath in FTR fading model\\
	$m_k$, $m_w$ &The fading severity parameter for the $k_{\rm th}$ user \& warden in FTR fading model\\
	$\Delta_k$, $\Delta_w$ & A parameter varying from $0$ to $1$ representing the similarity of two dominant waves for the $k_{\rm th}$ user \& warden in FTR fading model\\
	$\sigma_k$, $\sigma_w$ & The standard deviation of the diffuse received signal component for the $k_{\rm th}$ user \& warden in FTR fading model\\
	$ \upsilon_k$, $ \upsilon_w$ & The received average SNR for the $k_{\rm th}$ user \& warden in FTR fading model, and $ \upsilon =2 \sigma^{2}(1+K)$\\
	$m_{fk}$, $m_{fw}$ & The fading parameters for the $k_{\rm th}$ user \& warden in Fisher-Snedecor $\mathcal{F}$ fading model\\
	$m_{sk}$, $m_{sw}$ & The shadowing parameters for the $k_{\rm th}$ user \& warden in Fisher-Snedecor $\mathcal{F}$ fading model\\
	${\bar z}_k$, ${\bar z}_w$ & The average value of $\mathcal{F}$ random variables (RVs), i.e., $z$, for the $k_{\rm th}$ user \& warden in Fisher-Snedecor $\mathcal{F}$ fading model\\
	${}_1{F_1}\left( { \cdot ; \cdot ; \cdot } \right)$ & Confluent hypergeometric function \cite[eq. (9.210.1)]{gradshteyn2007}  \\
	$\Gamma \left( z \right) $ & Gamma function \cite[eq. (8.310.1)]{gradshteyn2007}  \\
	$\gamma\left(\cdot,\cdot\right)$ & Incomplete gamma function \cite[eq. (8.350.1)]{gradshteyn2007} \\
	${B\left( { \cdot , \cdot } \right)}$ & Beta function \cite[eq. (8.384.1)]{gradshteyn2007}  \\
	${}_2{F_1}\left( { \cdot , \cdot ; \cdot ; \cdot } \right)$ & Gauss hypergeometric function \cite[eq. (9.111)]{gradshteyn2007}  \\
	%	$G \, \substack{ m, n \\ p, q}(\cdot)$ & Meijer's $G$-function \cite[eq. (9.301)]{gradshteyn2007}  \\
	$H_{ \cdot  \cdot }^{ \cdot  \cdot }\left( { \cdot \left|  \cdot  \right.} \right)$ & Multivariate Fox's $H$-function \cite[eq. (A-1)]{mathai2009h}\\
	\hline
	\bottomrule
	\end{tabular}}
	\end{table}
	%=============================================================
	\section{System Model and SINR Analysis}\label{sec:FL}
	%=============================================================
	%=============================================================
	\subsection{System Description}	\label{sec:FL_fund} %checked
	%=============================================================
	\begin{figure}[h]
	\centering
	\includegraphics[scale=0.9]{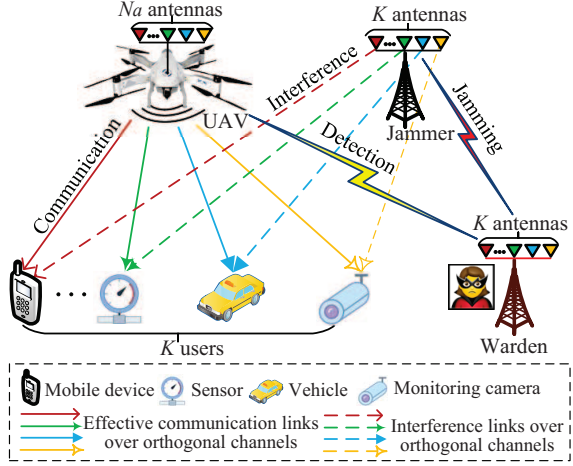}
	\caption{A jammer-aided multi-antenna UAV covert communication system.}
	\label {SystemMode}
	\end{figure}
	%=============================================================
	As shown in Fig. \ref{SystemMode}, we consider a jammer-aided multi-antenna UAV covert communication network. The UAV collects private data, such as surveillance images and users' travel trajectories, for $K$ users. Now, the UAV is transmitting the users' different data to the corresponding users simultaneously with the help of multi-antennas. Note that these users could be mobile devices, vehicles, or sensors. To prevent private data transmission from being detected by a malicious warden, a friendly jammer is used to improve the covertness of communication.
	
	Specifically, we assume that all users are equipped with a single antenna. Every user wants to improve its QoS, which includes data transmission and covertness, by purchasing UAV's transmitting power and jammer's jamming power. To avoid the interference among users, the UAV assigns different orthogonal frequency channels to use the $k_{\rm th}$ antenna to serve $k_{\rm th}$ user individually. Because the channel assignment information can be shared with the friendly jammer to facilitate a covert communication, the jammer uses its $k_{\rm th}$ antenna to jam the $k_{\rm th}$ channel to interfere the warden's detection. However, the jamming signals also have a negative impact on the user's SINR. We adopt a three-dimensional Cartesian coordinate system to represent the location. The locations of the user $U_k$, the jammer, the UAV, and the warden can be expressed by ${\mathbf{q}}_k = \left[ {{x_k},{y_k},{z_k}} \right]^T$ $\left( {\forall k \in \mathcal{K}} \right)$, $ {{\mathbf{q}}_j} = {\left[ {{x_j},{y_j},{z_j}} \right]^T}$, ${\mathbf{q}}_a = \left[ x_a,y_a,z_a  \right]^T $ and $ {\mathbf{q}}_w = {\left[ {{x_w},{y_w},,{z_w}} \right]^T}$, respectively. Therefore, the distances of the air-to-ground links, i.e., from UAV to the user $U_k$ and the warden, can be expressed as $D_{ak}$ and $D_{aw}$, respectively, where ${D}_{ai}={\|{\mathbf{q}}_a-{\mathbf{q}}_i\|}$ $(i=k,w)$. The distances of the ground-to-ground links from the jammer to the $k$th user and the warden are denoted by $D_{jk}$ and $D_{jw}$, respectively, where ${D}_{ji}={\|{\mathbf{q}}_j-{\mathbf{q}}_i\|}$.
	
	\begin{rem}
{\color{black} Because we use the generalized fading models with mathematical analysis complexity, our performance metrics are derived based on the scenario where one user is served by one antenna to reduce the complexity and obtain more insights. Note that our analysis can be generalized to the case of multi-antennas serving one user by a simple transformation \cite{zhang2020Performance}, i.e., the UAV uses $N_{k}$ antennas to serve the user $U_k$ and $ \sum\limits_{k = 1}^K {{N_k}}  = {N_a} $.} The reason is that we use FTR fading to model the small-scale fading in air-to-ground links (see Section \ref{ssaf21}), and the single FTR distribution can be regarded as the approximation to the distribution of the sum of FTR RVs \cite{zhang2020Performance}, i.e., if $ {N_k} \sim FTR\left( {{m_k},{K_k},{\sigma _k}^2,{\Delta _k}} \right) $, we have $ \sum\limits_{k = 1}^K {{N_k}}  \sim FTR\left( {\tilde m,\tilde K,{{\tilde \sigma }^2},\tilde \Delta } \right) $. The parameters $ \left\{ {\tilde m,\tilde K,{{\tilde \sigma }^2},\tilde \Delta } \right\} $ can be obtained easily \cite{zhang2020Performance}. A detailed analysis on transmit beamforming via the precoding associated with each user data, when $K_i$ antennas at UAV are used to serve one user, will be left for the future work.
	\end{rem}
	
	\subsection{Channel Model}
	Consider a large-scale and small-scale composite channel model. It is worth noting that air-to-ground and ground-to-ground connections could have various path loss exponents and thus require different fading models. The reason is that air-to-ground links have less obstruction, resulting in a much more stable line-of-sight (LoS) path. However, the ground environment in which the users, jammer, and warden are located, is typically dynamic, which may cause larger path loss exponents and composite fading and shadowing. In the following, we propose the path loss model and the small scale model for the air-to-ground links and the ground-to-ground links, respectively.
	\subsubsection{Path Loss Model}
	To simulate the air-to-ground and ground-to-ground links, different path loss coefficients are used, which are given as \cite{akdeniz2014millimeter}
	\begin{equation}
	L\left( d \right) = \left\{ {\begin{array}{*{20}{c}}
	{D_{ai}^{ - {\alpha _{ai}}},} \\ 
	{D_{ji}^{ - {\alpha _{ji}}},} 
	\end{array}} \right.\begin{array}{*{20}{l}}
	{{\text {for air-to-ground links,}}} \\ 
	{{\text {for ground-to-ground links,}}} 
	\end{array}
	\end{equation}
	where $\alpha_{ai}$ $(i \in \{k, w\})$ denotes the path loss exponents of the UAV-user $U_k$ link and the UAV-Warden link, respectively, and $\alpha_{ji}$ $(i=k,w)$ are the path loss exponents of the Jammer-user $U_k$ link and the Jammer-Warden link. Typical values of ${\alpha}$ can be defined in \cite[Table 1]{akdeniz2014millimeter}, while ${\alpha _{ai}} < {\alpha _{ji}}$ holds in general.
	
	\subsubsection{Small-Scale Model}\label{ssaf21}
	To unify system performance with various channel environments, generalized fading distributions are proposed, which include the most other fading distributions as special cases \cite{rabie2019full,du2020on}. In the following, we introduce the fluctuating two-ray (FTR) and Fisher-Snedecor $\mathcal{F}$ fading distributions to model the air-to-ground and ground-to-ground links, respectively. The reason is as follows:
	\begin{itemize}
	\item The FTR fading model \cite{romero2017fluctuating} is a generalization of the two-wave with diffuse power fading model, which allows the constant amplitude specular waves of LoS propagation to fluctuate randomly. Recent small-scale fading measurements of the $28$ {\rm GHz} outdoor millimeter-wave channels \cite{romero2017fluctuating} have shown that the FTR fading can provide a significantly better match to the real channel than the Rician fading. Therefore, we utilize the FTR fading model to illustrate the channel coefficients between the UAV and the user $U_k$ or warden, which have less obstruction, resulting in a much more stable LoS path.
	\item To create a realistic ground link model, we assume that each ground connection's small-scale fading follows Fisher-Snedecor $\mathcal{F}$ fading distribution, which has been proven to give a more thorough modeling and characterization of the simultaneous occurrence of multipath fading and shadowing \cite{yoo2017the}. In addition, the Fisher-Snedecor $\mathcal{F}$ model is more generic and covers several fading distributions as special cases, as well as being more mathematically tractable. For example, when the parameter $m_s$ tends to be infinity, the Fisher-Snedecor $\mathcal{F}$ distribution becomes the Nakagami-$m$ distribution. Moreover, special cases of Fisher-Snedecor $\mathcal{F}$ includes Rayleigh $(m = 1)$ and one-sided Gaussian $(m = 1/2)$. 
	\end{itemize}
	
	Let $ X\sim FTR\left( {m,K,{\sigma ^2},\Delta } \right) $. The PDF and CDF of the squared FTR RV $X$ can be expressed, respectively, as follows \cite{zhang2017new}:
	%$ X\sim FTR\left( {m,K,{\sigma ^2},\Delta } \right) $,
	{\small \begin{equation}\label{PDFFTR}
	{f_X}\!\left( x \right) = \frac{{{m^m}}}{{\Gamma \left( m \right)}}\sum\limits_{j = 0}^M  {\frac{{{K^j}{\alpha _j}}}{{j!}}} \frac{{{x^j}}}{{\Gamma \left( {j \!+ \!1} \right){{\left( {2{\sigma ^2}} \right)}^{j + 1}}}}\exp\!\left(\!{ - \frac{x}{{2{\sigma ^2}}}} \!\right),
	\end{equation}}	\noindent
	{\small \begin{equation}\label{CDFFTR}
	{F_X}\left( x \right) = \frac{{{m^m}}}{{\Gamma \left( m \right)}}\sum\limits_{j = 0}^M  {\frac{{{K^j}{\alpha _j}}}{{j!}}} \frac{1}{{\Gamma \left( {j + 1} \right)}}\gamma \left( {j + 1,\frac{x}{{2{\sigma ^2}}}} \right),
	\end{equation}}\noindent
	where
	{\small \begin{align}
	{\alpha _j}\! =& \sum\limits_{k = 0}^j \!{\left(\!\!\! {\begin{array}{*{20}{c}}
	j \\ 
	k 
	\end{array}}\!\!\! \right)} \sum\limits_{l = 0}^k \!{\left(\!\!\! {\begin{array}{*{20}{c}}
	k \\ 
	l 
	\end{array}} \!\!\!\right)} \Gamma \left( {j \!+\! m \!+\!2l\!-\! k} \right){\left( {m \!+\! K} \right)^{k - j - m - 2l}}\notag\\
	&\times{K^{2l - k}}{\left( {\frac{\Delta }{2}} \right)^{2l}}{\left( { - 1} \right)^{2l - k}}R_{j + m}^{k - 2l}\left( {{{\left( {\frac{{K\Delta }}{{m + K}}} \right)}^2}} \right),
	\end{align}}\noindent
	{\small \begin{equation}
	R_\upsilon ^\mu \!\left( x \right) \!\!=\!\! \left\{\!\!\! {\begin{array}{*{20}{c}}\!
	{\left(\! {\frac{{\upsilon  - \mu }}{2}}\! \right)\!\!\left(\! {\frac{{\upsilon  - \mu  + 1}}{2}} \!\right)\!\!\frac{{{x^\mu }}}{{\mu !}}{}_2{F_1}\!\!\left( {\frac{{\upsilon  + \mu }}{2}\!,\!\frac{{\upsilon  + \mu  + 1}}{2}\!;\!1\! + \!\mu ;x} \right)},{\mu  \in N,} \\ 
	{\frac{{{}_2{F_1}\!\left( {\frac{{\upsilon  - \mu }}{2},\frac{{\upsilon  - \mu  + 1}}{2};1 - \mu ;x} \right)}}{{\Gamma \left( {1 - \mu } \right)}}},{\text{otherwise},}
	\end{array}} \right.
	\end{equation}}\noindent
	and $M$ is a large constant that satisfies $ \sum\limits_{j = 0}^M {{\alpha _j}} \to 1 $. Typically, to obtain a satisfactory accuracy, e.g., $1-\sum\limits_{j = 0}^M {{\alpha _j}}<10^{-6}$, only less than $30$ terms are needed \cite{zheng2019sum}, which is easy to compute.
	
	Let $ Z\sim\mathcal{F}\left( {{m_f},{m_s},\bar z} \right) $, the PDF and CDF of the squared $\mathcal{F}$ RV $Z$ can be written as \cite{yoo2019comprehensive}
	{\small \begin{equation}
	{f_Z}\left( z \right) = \frac{{{m_f}^{{m_f}}{{\left( {{m_s} - 1} \right)}^{{m_s}}}{{\bar z}^{{m_s}}}{z^{{m_f} - 1}}}}{{B\left( {{m_f},{m_s}} \right){{\left( {{m_f}z + \left( {{m_s} - 1} \right)\bar z} \right)}^{{m_f} + {m_s}}}}},
	\end{equation}}\noindent
	and
	{\small \begin{align}
	{F_Z}\left( z \right) =& \frac{{{m_f}^{{m_f} - 1}{z^{{m_f}}}}}{{B\left( {{m_f},{m_s}} \right)\left( {{m_s} - 1} \right){^{{m_{f}}}}{{\bar z}^{{m_f}}}}}\notag\\&\times {}_2{F_1}\left( {{m_f},{m_f} + {m_s},{m_f} + 1; - \frac{{{m_f}z}}{{\left( {{m_s} - 1} \right)\bar z}}} \right).
	\end{align}}\noindent
	With the above accurate modeling of the channels, we can further analyze the SINR of the system and derive the PDF and CDF expressions in the following.
	\subsection{SINR Analysis}
	{\color{black}We consider $K$ users using different time-frequency resource blocks, which corresponds to the practical scenario when orthogonal multiple access is used \cite{huang2021jamming}.} $K$ antennas in UAV are used to serve $K$ users for tractable analysis under the generalized composite model of large-scale and small-scale fading channels. Therefore, the signal received by the user $U_k$ can be expressed as
	{\small \begin{equation}\label{signalex}
	{y_k} = \sqrt {D_{ak}^{ - {\alpha _{ak}}}{P_{ak}}} {h_{ak}}{s_k} + n + \sqrt {D_{jk}^{ - {\alpha _{jk}}}{P_{jk}}} {h_{jk}}, 
	\end{equation}}\noindent
	where ${P_{ak}}$ denotes the transmit power of the antenna which operates on the $k_{\rm th}$ channel, ${h_{ak}}$ is the channel coefficient of the air-to-ground link from UAV to user $U_k$, ${s_k}$ denotes the private data of user $U_k$, $ {\left\| {{s_k}} \right\|^2} = 1 $, ${P_{jk}}$ denotes the jamming power allocated to user $U_k$, ${h_{jk}}$ is the channel coefficient of the ground link form jammer to user $U_k$, and $n $ is the additive white Gaussian noise (AWGN) at the user with $n \sim {\cal C}{\cal N}\left( {0,{\kappa ^2}} \right)$. We denote the maximal sum transmit power of UAV by $P_{T}$. Hence, the transmit power constraint can be formulated as $\sum\limits_{k=1}^K P_{ak}\leq P_{T}$. Similarly, let $P_{J}$ denote the total jamming power, we have $\sum\limits_{k=1}^K P_{jk}\leq P_{J}$.
	
	With the help of \eqref{signalex}, the SINR of user $U_k$ can be expressed as
	{\small \begin{equation}
	{\gamma _k} = \frac{{D_{ak}^{ - {\alpha _{ak}}}{P_{ak}}h_{ak}^2}}{{{\kappa ^2} + D_{jk}^{ - {\alpha _{jk}}}{P_{jk}}h_{jk}^2}}\triangleq \frac{{{C_{1k}}{X_k}}}{{{\kappa ^2} + {C_{2k}}{Z_k}}},
	\end{equation}}\noindent
	where $X_k \triangleq h_{ak}^2 \sim FTR\left( {m_k,K_k,{\sigma ^2}_k,\Delta_k } \right)$, $ Z_k\triangleq h_{jk}^2 \sim\mathcal{F}\left( {{m_{fk}},{m_{sk}},\bar z_k} \right) $, $ {C_{1k}} \triangleq D_{ak}^{ - {\alpha _{ak}}}{P_{ak}} $, and $ {C_{2k}} \triangleq D_{jk}^{ - {\alpha _{jk}}}{P_{jk}} $.
	
	Next, we derive the PDF and CDF of ${\gamma _k}$ to further analyze the covert performance and provide a basis for the design of joint transmit power and jamming power allocation scheme.
	
	\section{Covert Performance Metrics and Analysis}\label{S3}
	\subsection{Statistical Expressions and Approximations}
	\subsubsection{Statistical Expressions}
	To perform the performance analysis of the system, the PDF and CDF expressions should be derived, which are presented in the following lemma.
	\begin{lemma}\label{Lemma-1}
	The PDF and CDF of the $k$-th user's SINR can be derived as \eqref{EQ-PDF} and \eqref{EQ-CDF}, respectively, shown at the bottom of the next page, where $ \Omega_k  \triangleq {\left( {{m_{sk}}\!-\!1} \right){{\bar z}_k}{C_{2k}}}$.
	\end{lemma}
	\begin{IEEEproof}
	Please refer to Appendix \ref{appendix1}.
	\end{IEEEproof}
	%----------------------------------------------------------------------------------------------------------------------------
	%----------------------------------------------------------------------------------------------------------------------------
	\newcounter{mycount}
	\begin{figure*}[b]
	\normalsize
	\setcounter{mycount}{\value{equation}}
	\hrulefill
	\vspace*{4pt}
	{\color{black} {\small\begin{align}\label{EQ-PDF}
	f_{\gamma_k}\!(\gamma)=& \!\frac{{{m_k}^{{m_k}}}\Gamma^{-1}\!\left( {{m_k}} \right)}{{\gamma \Gamma\!\left( {{m_{fk}}} \right)\Gamma\!\left( {{m_{sk}}} \right)}}{\left( {\frac{{{m_{fk}}{\kappa ^2}}}{{{\Omega _k}}}} \right)^{{m_{fk}}}}\sum\limits_{j = 0}^M {\frac{{{K_k}^j{\alpha _{kj}}}}{{\Gamma \left( {j + 1} \right)j!}}} H_{1,0:0,2:1,2}^{0,0:2,0:2,1}\!\!\left(\!\!\!\! {\left. {\begin{array}{*{20}{c}}
	{\frac{{\gamma {\kappa ^2}}}{{2{C_{1k}}\sigma _k^2}}}\\
	{\frac{{{\Omega _k}}}{{{m_{fk}}{\kappa ^2} - {\Omega _k}}}}
	\end{array}} \!\!\!\!\right|\!\!\!\!\begin{array}{*{20}{c}}
	{\left( {{m_{fk}} + {m_{sk}};1,1} \right): - :\left( {1,1} \right)}\\
	{ - \!:\!\left( {1 \!+\! j,\!1} \right)\left( {{m_{sk}},\!1} \right)\!:\!\left( {{m_{fk}},\!1} \right)\!\left( {{m_{sk}} \!+\! {m_{fk}},1} \right)}
	\end{array}} \!\!\!\right)
	\end{align}}\noindent
	{\small \begin{align}\label{EQ-CDF}
	F_{\gamma_k}\!(\gamma)\!= &\frac{{{m_k}^{{m_k}}}\Gamma^{-1}\!\left( {{m_k}} \right)}{{\Gamma\!\left( {{m_{fk}}} \right)\Gamma\!\left( {{m_{sk}}} \right)}}{\left(\! {\frac{{{m_{fk}}{\kappa ^2}}}{{{\Omega _k}}}} \!\right)^{{m_{fk}}}}\sum\limits_{j = 0}^M  {\frac{{{K_k}^j{\alpha _{kj}}}}{{\Gamma\!\left( {j \!+\! 1} \right)j!}}} H_{1,0:0,3:1,2}^{0,0:2,1:2,1}\!\!\left(\!\!\!\! {\left. {\begin{array}{*{20}{c}}
	{\frac{{\gamma {\kappa ^2}}}{{2{C_{1k}}\sigma _k^2}}}\\
	{\frac{{{\Omega _k}}}{{{m_{fk}}{\kappa ^2} - {\Omega _k}}}}
	\end{array}} \!\!\!\!\right|\!\!\!\!\begin{array}{*{20}{c}}
	{\left( {{m_{fk}} + {m_{sk}};1,1} \right):\left( {1,1} \right):\left( {1,1} \right)}\\
	{ - \!:\!\left( {1\! +\! j,\!1} \right)\!\left( {{m_{sk}},\!1} \right)\!\left( {0,\!1} \right)\!\!:\!\!\left( {{m_{fk}},\!1} \right)\!\left( {{m_{sk}} \!+\! {m_{fk}},\!1} \right)}
	\end{array}} \!\!\!\!\right)
	\end{align}}\noindent}
	\setcounter{equation}{\value{mycount}}
	\end{figure*}
	\addtocounter{equation}{2}
	%----------------------------------------------------------------------------------------------------------------------------
	%----------------------------------------------------------------------------------------------------------------------------
	
	{\color{black} In the following, we analyze truncation errors to demonstrate how infinite series affect the performance of the CDF expression. The truncation error of the ${F_{\gamma_k}}(\gamma)$ with respect to the first $M$ terms is given by \cite[eq. (32)]{du2021millimeter}
	{\small \begin{equation}\label{erroe}	
	\varepsilon \left( {{M}} \right) \buildrel \Delta \over = {F_Y}(\infty ) - {{\hat F}_Y}(\infty ).
	\end{equation}}\noindent
	Table \ref{tab1} illustrates the required truncation terms $M$ for different channel parameters to demonstrate the convergence of the series in \eqref{erroe}. We can observe that only less than $30$ terms are necessary for all considered cases with a desired accuracy of $10^{-5}$ or less.}
	\begin{table}[t]
	\caption{{\color{black} \label{tab1}Minimum Required Terms and Truncation Error for Different Parameters with ${K_k}=10$, $ {\Delta _k}=0.3$, $ {m_{sk}}=2$, $ {\bar z}=20$, $ D_{ak}=D_{jk}=10$, ${\alpha _{ak}}={\alpha _{jk}}=2$, $ {P_{jk}}=10$, and $ {\kappa ^2}=1$.}}
	\centering
	\begin{tabular}{|c|c|c|}
	\toprule
	\hline
	Parameter & {\small $\varepsilon \left( L \right)$} & {\small $M$}\\
	\hline
	{\small ${m_k}=4$, ${m_{fk}}=3$, ${P_{ak}}=13$, ${\sigma ^2}_k=0.2$} & {\small $7.34 \!\times\! 10^{-6}$} & {\small 21} \\
	\hline
	{\small ${m_k}=5$, ${m_{fk}}=5$, ${P_{ak}}=16$, ${\sigma ^2}_k=0.5$} & {\small $8.93\! \times \!10^{-6}$} & {\small 24} \\
	\hline
	{\small ${m_k}=7$, ${m_{fk}}=8$, ${P_{ak}}=26$, ${\sigma ^2}_k=0.3$} & {\small $8.58\! \times\! 10^{-6}$} & {\small 19} \\
	\hline
	{\small ${m_k}=2$, ${m_{fk}}=4$, ${P_{ak}}=14$, ${\sigma ^2}_k=0.7$} & {\small $7.02 \!\times\! 10^{-6}$} & {\small 22} \\
	\hline
	\bottomrule
	\end{tabular}
	\end{table}
	{\color{black} \subsubsection{Approximations Analysis}
	Although the previously derived analytical results have been obtained in closed-form, they may not provide many insights as to the factors affecting system performance. In the following, we analyze asymptotic performance that becomes tight in the high-SNDR or high-L regime.
	\begin{prop}\label{ApproProp1}
	When the jamming signal has weak interference to the receiver, i.e., the $k_{\rm th}$ user, the CDF of $\gamma_k$ can be approximated as
	{\small \begin{align}\label{afhlke}
	&F_{\gamma_k} ^{{\rm LJ}}\left( \gamma  \right)= \frac{{{m_k}^{{m_k}}}}{{\Gamma \left( {{m_k}} \right)}}{\left( {\frac{{{m_{fk}}{\kappa ^2}}}{{{m_{fk}}{\kappa ^2} - {\Omega _k}}}} \right)^{{m_{fk}}}}
	\notag\\&\times\!
	\sum\limits_{j = 0}^M \! {\frac{{{K_k}^j{\alpha _{kj}}}}{{\Gamma\!\left( {j \!+ \!2} \right)j!}}{{\left(\! {\frac{{{\kappa ^2}{\gamma }}}{{2\sigma _k^2{C_{1k}}}}}\!\right)}^{1 \!+\! j}}} {}_1{F_1}\!\left(\! {1 \!+\! j,2\! +\! j,\frac{{ - {\kappa ^2}{\gamma }}}{{2\sigma _k^2{C_{1k}}}}} \!\right).
	\end{align}}
\begin{IEEEproof}
	Please refer to Section \ref{NewAppen1} in Appendix \ref{NewAppen}.
\end{IEEEproof}
	\end{prop}
	\begin{prop}\label{ApproProp2}
	When the transmit power of UAV is high, the CDF of $\gamma_k$ can be approximated as \eqref{safg123}, shown at the bottom of this page. Note that when $m_{sk}$ is not an integer, $j_s$ is an integer satisfying the condition $ {j_s} + 1 < {m_{sk}} < {j_s} + 2 $, and $ \Xi \left( j \right) \triangleq \Gamma \left( {{m_{sk}} - 1 - j} \right){K_k}^j{\alpha _{kj}} $ for all $j$. When $m_{sk}$ is an integer, we set $ {j_s} = {m_{sk}} - 1 $, $ \Xi \left( {{j_s}} \right) \triangleq {K_k}^{{j_s}}{\alpha _{k{j_s}}} $, and $ \Xi \left( j \right) \triangleq \Gamma \left( {{m_{sk}} - 1 - j} \right){K_k}^j{\alpha _{kj}} $ for $j \ne {j_s}$.
	%----------------------------------------------------------------------------------------------------------------------------
%----------------------------------------------------------------------------------------------------------------------------
\newcounter{mycount543}
\begin{figure*}[b]
	\normalsize
	\setcounter{mycount543}{\value{equation}}
	\hrulefill
	\vspace*{4pt}
	{\color{black} {\small \begin{align}\label{safg123}
				F_{{\gamma _k}}^{\rm HT}\left( \gamma  \right)& =\frac{{{m_k}^{{m_k}}}}{{{m_{sk}}\Gamma\!\left( {{m_k}} \right)B\!\left( {{m_{sk}},\!{m_{fk}}} \right)}}{\left(\! {\frac{{{\Omega _k}}}{{{m_{fk}}}}\frac{1}{{2\sigma _k^2}}\frac{\gamma }{{{C_{1k}}}}} \!\right)^{{m_{sk}}}}\!\!\sum\limits_{j = {j_s} + 1}^M {\frac{{\Gamma\! \left(\! {1\! +\! j\! -\! {m_{sk}}} \!\right){K_k}^j{\alpha _{kj}}}}{{\Gamma \!\left( {j \!+ \!1} \right)j!}}} + \frac{{{m_k}^{{m_k}}}}{{\Gamma\!\left( {{m_k}} \right)\!\Gamma\!\left( {{m_{sk}}} \right)}}{\left(\! {\frac{{{m_{fk}}{\kappa ^2}}}{{{\Omega _k}}}} \!\right)^{{m_{fk}}}}
				\notag\\&\times
				\sum\limits_{j = 0}^{{j_s}} {\frac{\Xi \left( j \right)}{{B\!\left( \!{{m_{fk}} \!+ \!{m_{sk}} \!-\! j \!- \!1,j \!+ \!1} \!\right)\left( {j \!+\! 1} \right)!}}} {\left(\! {\frac{{{\kappa ^2}\gamma }}{{2\sigma _k^2{C_{1k}}}}} \!\right)^{j + 1}}{}_2{F_1}\!\left(\! {{m_{fk}},{m_{sk}} \!+\! {m_{fk}};{m_{fk}}\! +\! {m_{sk}} \!- \!j \!- \!1;\frac{{{\Omega _k} \!- \!{m_{fk}}{\kappa ^2}}}{{{\Omega _k}}}} \!\right)
		\end{align}}\noindent}
	\setcounter{equation}{\value{mycount543}}
\end{figure*}
\addtocounter{equation}{1}
%----------------------------------------------------------------------------------------------------------------------------
%----------------------------------------------------------------------------------------------------------------------------
\begin{IEEEproof}
	Please refer to Section \ref{NewAppen2} in Appendix \ref{NewAppen}.
\end{IEEEproof}
	\end{prop}
\begin{prop}\label{ApproProp3}
	When the transmit power of UAV is high and the interference from the jamming signal is low, the CDF of $\gamma_k$ can be approximated as
	{\small \begin{align}\label{1dsad}
			F_{{\gamma _k}}^{{\rm HTLJ}}\left( \gamma  \right) = &\frac{{{m_k}^{{m_k}}}}{{\Gamma \left( {{m_k}} \right)}}{\left( {\frac{{{m_{fk}}{\kappa ^2}}}{{{m_{fk}}{\kappa ^2} - {\Omega _k}}}} \right)^{{m_{fk}}}}
			\notag\\&\times
			\sum\limits_{j = 0}^M  {\frac{{{K_k}^j{\alpha _{kj}}}}{{\Gamma \left( {j + 1} \right)\left( {1 + j} \right)!}}} {\left( {\frac{{{\kappa ^2}{\gamma}}}{{2\sigma _k^2{C_{1k}}}}} \right)^{1 + j}}.
	\end{align}}
\begin{IEEEproof}
	Please refer to Section \ref{NewAppen3} in Appendix \ref{NewAppen} .
\end{IEEEproof}
\end{prop}
	\begin{figure}[t]
		\centering
		\includegraphics[scale=0.5]{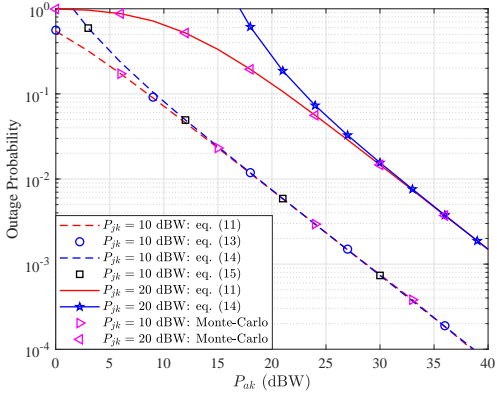}	
		\caption{{\color{black} The outage probability versus the transmit power, with $m_k=2$, $K_k=2$, $2\sigma_k^2(1+K_k)=15$ ${\rm dB}$, $\Delta_k=0.7$, $ {{\kappa ^2}}=1$ ${\rm dB}$, $ {{m_{fk}}}=3$, $ {{m_{sk}}} =3$, $\bar z_k =1$, $\gamma_{\rm th}=1$, $D_{jk}=10$ ${\rm m}$, $D_{ak}=5$ ${\rm m}$, $\alpha_{jk}=3$, $\alpha_{ak}=2$, and different values of jamming power.}}
		\label{Appro}
	\end{figure}

	With the help of \eqref{afhlke} and \eqref{1dsad}, the derived performance metrics can be simplified. For example, the outage probability (OP) is defined as the probability that the received SNR falls below a given outage threshold $\gamma _{\rm th}$, which means that $P_{\rm out} = {\mathbb P} \left( \gamma < \gamma _{\rm th}\right) =F_{\gamma}\left( \gamma _{\rm th}\right) $. The OP of the jammer-aided UAV communication system can be directly evaluated by using \eqref{EQ-CDF}. When the interference from the jamming signal is small or the transmit power is large or both conditions are satisfied, we can use \eqref{afhlke}, \eqref{safg123}, and \eqref{1dsad} to obtain the corresponding approximate OP expressions, respectively. Figure \ref{Appro} depicts the OP performance versus the transmit power. We can observe that a good agreement exists between the analytical ({\it red lines}) and Monte-Carlo simulation results ({\it triangles}), which validates the proposed analytical expressions. In the high-transmit power regime, the approximate values obtained by \eqref{safg123} and \eqref{1dsad} ({\it black lines and black squares, respectively}), match well the exact ones obtained by \eqref{EQ-CDF}. Furthermore, when the jamming power is relatively low, i.e., $P_{jk}=10$ ${\rm dBW}$, the calculation results of \eqref{afhlke} ({\it black circles}) are always close to the results of \eqref{EQ-CDF} under any transmit power. 
	}
	\subsection{Covert Performance Metrics}
	{\color{black} Then we analyze the detection behavior of warden.} The warden has a binary choice between the null hypothesis, $\mathcal{H}_0$, that UAV is silent, and the alternate hypothesis, $\mathcal{H}_1$, that UAV is transmitting. Moreover, the warden can perform statistical hypothesis testing based on the received average power which includes the noise power and jamming power in $\mathcal{H}_0$ and additionally includes the received signal power in $\mathcal{H}_1$. The received signals at the warden's $k_{\rm th}$ antenna can be expressed as
	{\small \begin{equation}
	{y_{wk}} = \left\{ {\begin{array}{*{20}{l}}
	{{\kappa ^2} + D_{jw}^{ - {\alpha _{jw}}}{P_{jk}}h_{jw}^2}, \quad\qquad\qquad\qquad \mathcal{H}_0,	 \\ 
	{D_{aw}^{ - {\alpha _{aw}}}{P_{ak}}h_{aw}^2 + {\kappa ^2} + D_{jw}^{ - {\alpha _{jw}}}{P_{jk}}h_{jw}^2}, \: \mathcal{H}_1.
	\end{array}} \right.
	\end{equation}}\noindent
	Let $\mathcal{D}_1$ and $\mathcal{D}_0$ represent the warden's decisions in favor of $\mathcal{H}_1$ and $\mathcal{H}_0$, respectively. The warden's decisions are based on a threshold-based rule, which is commonly adopted \cite{jiang2021covert,zhou2019joint,soltani2018covert} and advocates $\mathcal{D}_1$ and $\mathcal{D}_0$ when the received power is larger and not larger than a predefined threshold, respectively. We can observe that erroneous decision occurs in two cases: Warden sides with $\mathcal{D}_1$ when the $\mathcal{H}_0$ is true, which is called {\it false alarm}, and warden sides with $\mathcal{D}_0$ when the $\mathcal{H}_1$ is true, which is called {\it missed detection}. The error probability is defined as the likelihood of the warden making an incorrect decision. Note that the covertness of communication is guaranteed if the warden's detection error for each user is always larger than a threshold, $\xi_{\rm th}$, which is arbitrarily close to $1$. Under such conditions, the communication rate is regarded as the covert rate. We use the detection error probability and covert rate to assess the system's performance in the following.
	\subsubsection{Detection Error Probability}
	The performance of warden's hypothesis test can be measured by the detection error probability, which is defined as
	{\small {\color{black} \begin{align}
	{\xi _k} =& {{\mathbb P}_{FA}} + {{\mathbb P}_{MD}} = \Pr \left( {{\kappa ^2} + D_{jw}^{ - {\alpha _{jw}}}{P_{jk}}h_{jw}^2 > {\varepsilon _k}} \right)\notag\\&
	+ \Pr \left( {D_{aw}^{ - {\alpha _{aw}}}{P_{ak}}h_{aw}^2 + {\kappa ^2} + D_{jw}^{ - {\alpha _{jw}}}{P_{jk}}h_{jw}^2 < {\varepsilon _k}} \right),
	\end{align}}}\noindent
	where $\mathbb{P}_{FA}=\mathbb{P}(\mathcal{D}_1|\mathcal{H}_0)$ denotes the false alarm probability, and $\mathbb{P}_{MD}=\mathbb{P}(\mathcal{D}_0|\mathcal{H}_1)$ denotes the miss detection probability. 
	
	We assume that the warden can set different detection thresholds for different users to minimize the corresponding detection error probability. Let ${\varepsilon _{k}}$ denote the detection threshold for user $U_k$. The detection error probability can be obtained in Theorem \ref{IEEEPROOF3}.
	
	\begin{theorem}\label{IEEEPROOF3}
	The detection error probability is derived as \eqref{detection}, shown at the bottom of this page, where $C_{1w}\triangleq D_{aw}^{ - {\alpha _{aw}}}{P_{ak}}$, $C_{2w}\triangleq D_{jw}^{ - {\alpha _{jw}}}{P_{jk}}$, and $ \Omega_w  \triangleq {\left( {{m_{sw}}\!-\!1} \right){{\bar z}_w}{C_{2w}}}$.
	\end{theorem}
	\begin{IEEEproof}
	Please refer to Appendix \ref{appendix3}.
	\end{IEEEproof}
	%----------------------------------------------------------------------------------------------------------------------------
	%----------------------------------------------------------------------------------------------------------------------------
	\newcounter{mycount2}
	\begin{figure*}[b]
	\normalsize
	\setcounter{mycount2}{\value{equation}}
	\hrulefill
	\vspace*{4pt}
	{\small 	{\color{black} \begin{align}\label{detection}
	\xi_k =& 1 - \frac{{{m_{fw}}^{{m_{fw}} - 1}{{\left( {{\varepsilon _{k}} - {\kappa ^2}} \right)}^{{m_{fw}}}}}}{{{C_{2w}}^{{m_{fw}}}B\!\left( {{m_{fw}},{m_{sw}}} \right)\left( {{m_{sw}} - 1} \right){^{{m_{fw}}}}{{\bar z}_w}^{{m_{fw}}}}}{}_2{F_1}\!\left( {{m_{fw}},{m_{fw}} \!+ \!{m_{sw}},{m_{fw}} \!+\! 1; - \frac{{{m_{fw}}\left( {{\varepsilon _{k}} \!- \!{\kappa ^2}} \right)}}{{\Omega_w}}} \right)\notag\\
	&+\frac{{{m_w}^{{m_w}}}}{{\Gamma \left( {{m_{fw}}} \right)\Gamma \left( {{m_{sw}}} \right)\Gamma \left( {{m_w}} \right)}}\sum\limits_{j = 0}^\infty  {\frac{{{K_w}^j{\alpha _{wj}}}}{{j!\Gamma \left( {j + 1} \right)}}} H_{1,0:1,3:1,1}^{0,0:3,0:1,1}\left( {\left. {\begin{array}{*{20}{c}}
	{\frac{{{\Omega _w}}}{{{m_{fw}}\left( {{\varepsilon _{th}} - {\kappa ^2}} \right)}}}\\
	{\frac{{2{C_{1w}}\sigma _w^2}}{{{\varepsilon _{th}} - {\kappa ^2}}}}
	\end{array}} \right|\begin{array}{*{20}{c}}
	{\left( {1;1,1} \right):\left( {1 - {m_{fw}},1} \right)\left( {1,1} \right):\left( { - j,1} \right)}\\
	{ - :\left( {0,1} \right)\left( {{m_{sw}},1} \right)\left( {1,1} \right):\left( {0,1} \right)}
	\end{array}} \right)
	\end{align}}}\noindent
	\setcounter{equation}{\value{mycount2}}
	\end{figure*}
	\addtocounter{equation}{1}
	%----------------------------------------------------------------------------------------------------------------------------
	%----------------------------------------------------------------------------------------------------------------------------
	
	From \eqref{detection}, we can observe that the UAV and jammer can improve the covertness of communication by jointly adjusting $P_{ak}$ and $P_{jk}$ assigned to user $U_k$. However, the warden can obtain the theoretical minimum error probability by adjusting ${\varepsilon _{k}}$. {\color{black}To demonstrate the robustness of the optimal allocation algorithm, the worst-case scenario is considered where the warden can obtain the perfect CSI needed to optimize its detection threshold \cite{hu2018covert}. The warden's algorithm for optimizing the threshold is introduced in Section \ref{section4A}. If our power allocation algorithm can still guarantee that the detection error probability is arbitrarily close to $1$ under the worst-case scenario, covert communication is successfully achieved.}
	
	{\color{black} \begin{rem}
			Due to channel estimation errors in practical systems, the warden may have imperfect CSI of the users. In particular, the warden has an estimated version of the channels. According to the worst-case method, channel mismatches occur in the bound set, where the upper bounds are known constants \cite{forouzesh2020joint}, i.e., $ {{e_{aw}}} $ and $ {{e_{jw}}} $ are the estimation errors of the UAV-warden and jammer-warden links, respectively. Thus, the DEP achieved by the warden based on imperfect CSI can be expressed as ${\xi_k}'$ $=$ $\Pr\!\left(\! {D_{aw}^{{\alpha _{aw}}}{P_{ak}}\left( {h_{aw}\!+ \!{e_{aw}}} \right)^2 \! +\! {\kappa ^2}\!+ \!D_{jw}^{{\alpha _{jw}}}{P_{jk}}\left( {h_{jw}\! + \!{e_{jw}}} \right)^2\! < \!{\varepsilon _k}} \!\right)$ $+$ $\Pr \left( {{\kappa ^2} + D_{jw}^{{\alpha _{jw}}}{P_{jk}}\left( {h_{jw} + {e_{jw}}} \right)^2 > {\varepsilon _k}} \right)$. Because the PDF and CDF of both $h_{aw}$ and $h_{jw} $ are known, we can obtain the statistical expressions of $\left( {h_{aw}+ {e_{aw}}} \right)$ and $\left( {h_{jw}+ {e_{jw}}} \right)$ easily. Then, by following the same derivation methods in Appendix \ref{appendix3}, the closed-form expression of DEP with imperfect CSI can be obtained. Note that if the warden uses the expression of DEP with imperfect CSI to determine the threshold, the warden will inaccurately select the detection threshold that does not achieve the minimal DEP, as shown in Fig. \ref{DEP}.
	\end{rem}}
	{\color{black} \begin{rem}
			If we consider that the jammer uses multiple antennas, i.e., $L_J$ antennas, to assist one user, we can design the jamming signal beamforming vector to achieve better covertness. Let ${\bf {h_{jw}}} = \left[ {{h_{jw,1}},{h_{jw,2}}, \ldots ,{h_{jw,{L_J}}}} \right]$ denote the MISO channel fading vector between the jammer and the warden, and ${\bf {v_k}} = \left[ {{v_{k,1}},{v_{k,2}}, \ldots ,{v_{k,{L_J}}}} \right]$ denote the jamming signal vector, where {\small $ \left\| {{{\bf{v}}_{\bf{k}}}} \right\| \le 1 $}. Then, the received jamming signal can be expressed as $ {y_{jw}}$ $=$ $\sqrt {D_{jw}^{ - {\alpha _{jw}}}}{\bf {h_{jw}}}{\bf {v_k}}^T$. Using the maximum ratio transmitting \cite{lo1999maximum,tokgoz2006performance}, we can define ${\bf {v_k}}$ as ${{\bf{v}}_{\bf{k}}} = {{{\bf{h}}_{{\bf{jw}}}}}/{{\left\| {{{\bf{h}}_{{\bf{jw}}}}} \right\|}}$, and the corresponding DEP is re-written as {\small $ {\xi _{k}^{{\small (MRT)}}}$  $=$ $\Pr \!\left(\! {D_{aw}^{{\alpha _{aw}}}\!{P_{ak}}{{\left(\! {{h_{aw}} \!+ \!{e_{aw}}} \!\right)}^2} \!+\! {\kappa ^2} \!+\! {P_{jk}}D_{jw}^{{\alpha _{jw}}}{{\left(\! {\sum\limits_{\ell  = 1}^{{L_J}} {{h_{jw,\ell }}} } \!\right)}^2}\! <\! {\varepsilon _k}} \right)$ $+$ $\Pr \!\left(\! {{\kappa ^2} \!+ \! {P_{jk}}D_{jw}^{{\alpha _{jw}}}{{\left(\! {\sum\limits_{\ell  = 1}^{{L_J}} {{h_{jw,\ell }}} } \!\right)}^2} \!> \!{\varepsilon _k}} \!\right) $}. Because ${h_{jw,\ell }}$ is the Fisher-Snedecor $\mathcal{F}$ RV, the distribution of the sum of $L_J$ $\mathcal{F}$ RVs, i.e., $ {\sum\limits_{\ell  = 1}^{{L_J}} {{h_{jw,\ell }}} } $, can be approximated by the single Fisher-Snedecor $\mathcal{F}$ fading distribution \cite{du2020sum}. Thus, we can analyze the DEP following similar methods in Theorem \ref{IEEEPROOF3}.
	\end{rem}}
%	The covert rate is derived in the following.
	\subsubsection{Covert Communication Rate}
	Using the PDF and CDF expressions of $\gamma_k$ in Lemma \ref{Lemma-1}, we can obtain the covert rate of the user $U_k$.
	\begin{theorem}\label{IEEEPROOF2}
	In our considered system, the covert rate of the $k$-th user is given by \eqref{EQ-1}, shown at the bottom of the next page.
	%	where $ \Omega_k  \triangleq {\left( {{m_{sk}}\!-\!1} \right){{\bar z}_k}{C_{2k}}}$.
	\end{theorem}
	\begin{IEEEproof}
	Please refer to Appendix \ref{appendix2}.
	\end{IEEEproof}
	%----------------------------------------------------------------------------------------------------------------------------
	%----------------------------------------------------------------------------------------------------------------------------
	\newcounter{mycount3}
	\begin{figure*}[b]
	\normalsize
	\setcounter{mycount3}{\value{equation}}
	\hrulefill
	\vspace*{4pt}
	{\small {\color{black} \begin{align}\label{EQ-1}
	R_k\!=&\frac{{ - {m_k}^{{m_k}}}\Gamma^{-1} \left( {{m_k}} \right)}{{\ln 2\Gamma\!\left( {{m_{fk}}} \right)\!\Gamma\!\left( {{m_{sk}}} \right)}}{\left( {\frac{{{m_{fk}}{\kappa ^2}}}{{{\Omega _k}}}} \right)^{{m_{fk}}}}\sum\limits_{j = 0}^\infty  {\frac{{{K_k}^j{\alpha _{kj}}}}{{\Gamma\!\left( {j\! + \!1} \right)j!}}}\!H_{1,0:1,3:1,2}^{0,0:3,1:2,1}\!\!\left(\!\!\!\! {\left. {\begin{array}{*{20}{c}}
	{\frac{{{\kappa ^2}}}{{2{C_{1k}}\sigma _k^2}}}\\
	{\frac{{{\Omega _k}}}{{{m_{fk}}{\kappa ^2} - {\Omega _k}}}}
	\end{array}} \!\!\!\!\right|\!\!\!\!\begin{array}{*{20}{c}}
	{\left( {{m_{fk}} + {m_{sk}};1,1} \right):\left( {1,1} \right):\left( {1,1} \right)}\\
	{ - \!:\!\left( {1 \!+ \!j,1} \right)\!\left( {{m_{sk}},\!1} \right)\!\left( {0,\!1} \right)\!:\!\left( {{m_{fk}},\!1} \right)\!\left( {{m_{sk}} \!+ \!{m_{fk}},\!1} \right)}
	\end{array}} \!\!\!\!\right)
	\end{align}}}\noindent
	\setcounter{equation}{\value{mycount3}}
	\end{figure*}
	\addtocounter{equation}{1}
	%----------------------------------------------------------------------------------------------------------------------------
	%----------------------------------------------------------------------------------------------------------------------------
	
	While ensuring that the communication is covert, every user wants to communicate at a rate as high as possible. From \eqref{EQ-1}, we can observe that, except for the unchangeable channel coefficients, the maximum covert rate for user $U_k$ can be achieved by adjusting $P_{ak}$ and $P_{jk}$. To allocate the limited resources fairly to multiple users, we next formulate the problem with the help of game theory by regarding users as players in a bargaining game.
	
	\section{Problem Formulation}\label{section4}
	The user's QoS requirements consist of two categories, which are the covertness of communication and the transmission performance. {\color{black}To achieve covert communication, each user asks for an adequate amount of jammer power. Additionally, each user wants the UAV to assign an adequate amount of transmit power to improve the covert rate.}
	However, there are two key challenges: 
	\begin{itemize}
	\item If too much transmit power is provided to a single user, the communication's covertness may be compromised, as warden can detect the transmission more accurately. Furthermore, because the jamming and the data signals are in the same frequency band, too high jamming power will reduce the covert rate. As a result, we have to optimize the allocation of transmit and jamming power.
	\item There are $K$ users being served at the same time, but both the transmit power of the UAV and jamming power of the jammer are restricted. Limited resources lead to gaming among users. Thus, we may use the bargaining game theory to solve the power allocation problem. The users can be viewed as the players in the game. 
	\end{itemize}
	To tackle the aforementioned two key challenges, we represent the optimization problem as a bargaining game to maximize the user's covert rate with limited transmit and jamming power, while ensuring the covertness. To demonstrate the robustness, the game is modeled under the warden's optimal detection threshold, which represent the worst-case scenario for the users. In the following, we introduce the method for the warden to obtain the optimal threshold, and then propose the bargaining game model.
	\subsection{Optimal Detection Threshold at Warden}\label{section4A}
	\subsubsection{Converge Analysis}\label{039falef}
\begin{figure}[t]
	\centering
	\includegraphics[scale=0.5]{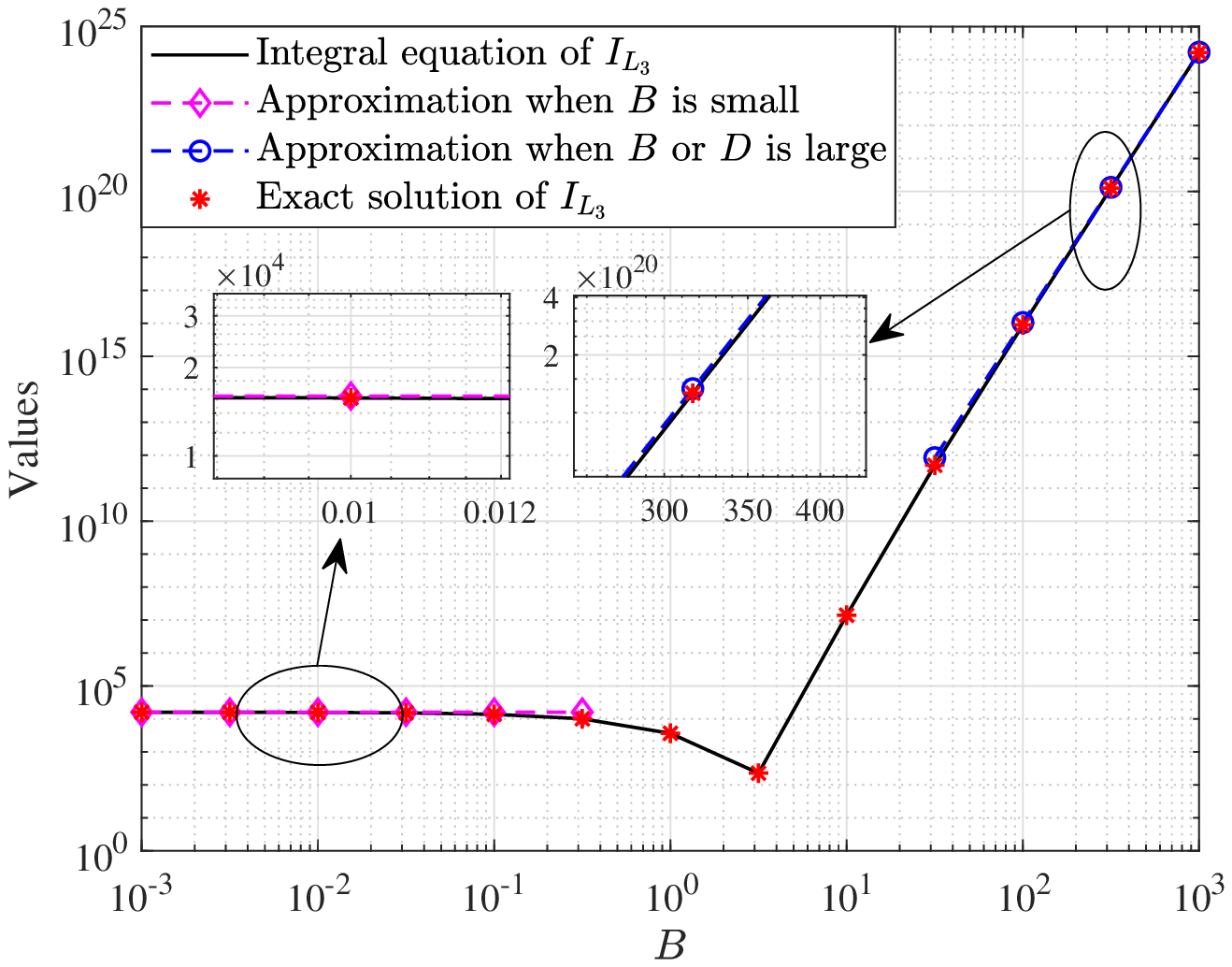}
	\caption{{\color{black} Equation real-part values on the left and right-hand sides of $I_{L_3}$ versus $B$, with $A=3.2$, $C=8.2$, $D=2$.}}
	\label{Math}
	\end{figure}

	{\color{black} From \eqref{detection}, we can see that the detection threshold, ${\varepsilon _{k}}$, transmit and jamming powers exist in both Gauss hypergeometric function and multivariate Fox's $H$-function, which means that it is difficult to analyze whether \eqref{detection} is convex directly. Therefore, we first study the following integral equation.
	\begin{lemma}\label{appendixDD}
	The $I_L\triangleq \int_{{T_1}}^{{T_2}} {{t^A}{{\left( {B - t} \right)}^C}{e^{ - Dt}}{\rm d}t}$ with different cases of $T_1$ and $T_2$ can be derived as \eqref{intr}, shown at the bottom of the next page, where $U$ is the tricomi confluent hypergeometric function \cite{buchholz2013confluent}.
	\end{lemma}
	\begin{IEEEproof}
	Please refer to Appendix \ref{appendixD}.
	\end{IEEEproof}
	To the best of our knowledge, an exact solution of $I_L$ cannot be found in any function books, e.g., \cite{Prudnikov1986Integrals,gradshteyn2007,web}. For the third case, i.e., when $\left( T_1=0, T_2=\infty \right) $, due to the complexity of the tricomi confluent hypergeometric function, it may be difficult to be used for convergent analysis. Thus, we derive two kinds of approximate results. As shown in Fig. \ref{Math}, the approximation expressions that we obtain are highly accurate. For the first two cases, the approximate expressions can be obtained following the similar methods in Proposition~\ref{ApproProp2}.
	
	Because that the exponential function is common in many channel fading models \cite{romero2017fluctuating,zhang2017new,yoo2017the}, Lemma \ref{appendixDD} is useful in the performance analysis of wireless communications systems. Using Lemma \ref{appendixDD}, we can analyze \eqref{detection} to prove its convexity in the following theorem.}
	%----------------------------------------------------------------------------------------------------------------------------
	%----------------------------------------------------------------------------------------------------------------------------
	\newcounter{mycount523}
	\begin{figure*}[b]
	\normalsize
	\setcounter{mycount523}{\value{equation}}
	\hrulefill
	\vspace*{4pt}
	{\small {\color{black} \begin{align}\label{intr}
	&\int_{{T_1}}^{{T_2}} {{t^A}{{\left( {B - t} \right)}^C}{e^{ - Dt}}dt}  =
	\notag\\&\!
	\!	\left\{\!\!\! {\begin{array}{*{20}{l}}
	{{\rm{Cases}}}&{{\rm{Results}}}&{{\rm{Conditions}}}\\\!\!\!
	\begin{array}{l}
	{T_1} \!=\! 0,\\
	{T_2}\! =\! T
	\end{array}&\!\!{{I_{{L_1}}} = \frac{{{B^C}{T^{1 + A}}}}{{\Gamma \left( { - C} \right)}}H_{1,1:0,1;1,1}^{0,1:1,0;1,1}\!\!\left(\!\!\!\! {\left. {\begin{array}{*{20}{c}}
	{DT}\\
	{ - {B^{ - 1}}T}
	\end{array}} \!\!\!\right|\!\!\!\begin{array}{*{20}{c}}
	{\left( { - A;1,1} \right): - :\left( {1 + C,1} \right)}\\
	{\left( { - 1 - A;1,1} \right)\!:\!\left( {0,\!1} \right)\!:\!\left( {0,1} \right)}
	\end{array}} \right)}&\!\!\!\!\!\begin{array}{l}
	\left\{ {{\mathop{\rm Re}\nolimits} \left\{ {\frac{B}{T}} \right\} > 1||{\mathop{\rm Re}\nolimits} \left\{ {\frac{B}{T}} \right\} \le 0||\frac{B}{T} \notin {\mathop{\rm Re}\nolimits} } \right\}\\
	\& {\mathop{\rm Re}\nolimits} \left\{ D \right\} > 0\& C \ne 0, - 1, - 2, \ldots 
	\end{array}\\\!\!\!
	\begin{array}{l}
	{T_1} \!=\! T,\\
	{T_2} \!=\! \infty 
	\end{array}&\!\!\!\!\!\begin{array}{l}
	{I_{{L_2}}} = {\left( { - 1} \right)^C}{\left( {{T_1}} \right)^{1 + A + C}}\\
	\times\! H_{2,1:2,0;0,1}^{0,2:0,1;1,0}\!\!\left(\!\!\!\!\! {\left. {\begin{array}{*{20}{c}}
	{{{\left( {D{T_1}} \right)}^{ - 1}}}\\
	{ - B{T^{ - 1}}}
	\end{array}} \!\!\!\right|\!\!\!\!\begin{array}{*{20}{c}}
	{\left( {2 \!+\! A \!+\! C;\!1,\!1} \right)\!\left( { - \!A, - \!1,\!1} \right)\!:\!\left( {1,\!1} \right)\!\left( {1 \!+ \!A,1} \right)\!:\!- }\\
	{\left( {1 + A + C;1,1} \right): - :\left( {0,1} \right)}
	\end{array}}\!\!\!\! \right)
	\end{array}&\!\!\!\!\!\begin{array}{l}
	{\mathop{\rm Re}\nolimits} \left\{ B \right\} > 0\& {\mathop{\rm Re}\nolimits} \left\{ D \right\} > 0\\
	\& {\mathop{\rm Re}\nolimits} \left\{ T \right\} > 0\& {\rm Im}\left\{ T \right\} = 0
	\end{array}\\\!\!\!
	\begin{array}{l}
	{T_1} \!=\! 0,\\
	{T_2} \!=\! \infty 
	\end{array}&\!\!\!\!\!\begin{array}{l}
	{I_{{L_3}}} = \Gamma \left( {1 + A} \right){B^C}{\left( { - B} \right)^{1 + A}}U\left( {1 + A,2 + A + C, - BD} \right)\\
	\qquad \approx \Gamma \left( {1 + A} \right){B^C}{D^{ - A - 1}}, ({\rm when} \: B\: {\rm or}\: D {\rm \: is\: large}) \\
	\qquad \approx \frac{{\Gamma \left( { - 1 - A - C} \right)}}{{\Gamma \left( { - C} \right)}} + \frac{{\Gamma \left( {1 + A + C} \right)}}{{\Gamma \left( {1 + A} \right){{\left( { - BD} \right)}^{1 + A + C}}}}, ({\rm when} \: B\: {\rm or}\: D {\rm \: is\: small)}
	\end{array}&\!\!\!\!\!\begin{array}{l}
	{\mathop{\rm Re}\nolimits} \left\{ A \right\} >  - 1\& {\mathop{\rm Re}\nolimits} \left\{ B \right\} > 0\\
	\& {\mathop{\rm Re}\nolimits} \left\{ D \right\} > 0
	\end{array}\\\!\!\!
	\begin{array}{l}
	{T_1} \!=\! 0,\\
	{T_2} \!=\! B
	\end{array}&\!\!{{I_{{L_4}}} = \frac{{{B^{1\! +\! A \!+\! C}}\Gamma\! \left( {1 \!+\! A} \right)\Gamma\! \left( {1\! + \!C} \right)}}{{\Gamma \left( {2 + A + C} \right)}}{}_2{F_1}\left( {1 + A,2 + A + C, - BD} \right)}&\!\!\!\!\!\begin{array}{l}
	{\mathop{\rm Re}\nolimits} \left\{ A \right\} >  - 1\& {\mathop{\rm Re}\nolimits} \left\{ C \right\} >  - 1\\
	\& {\mathop{\rm Re}\nolimits} \left\{ B \right\} > 0\& {\mathop{\rm Im}\nolimits} \left\{ B \right\} = 0
	\end{array}
	\end{array}} \right.
	\end{align}}}\noindent
	\setcounter{equation}{\value{mycount523}}
	\end{figure*}
	\addtocounter{equation}{1}
	%----------------------------------------------------------------------------------------------------------------------------
	%----------------------------------------------------------------------------------------------------------------------------
	
	\begin{theorem}\label{appendixEE}
	Under a given allocation scheme of transmit and jamming power, the detection error probability is convex with respect to the detection threshold, ${\varepsilon_{k}}$.
	\end{theorem}
	\begin{IEEEproof}
	Please refer to Appendix \ref{appendixE}.
	\end{IEEEproof}
	Therefore, the optimal detection threshold, ${\varepsilon^{opt} _{k}}$, can be derived by solving $ \frac{{\partial {\xi _k}}}{{\partial {\varepsilon _k}}} = 0 $. However, in many cases, the warden cannot obtain the perfect channel state information. Thus, in the following, we propose a PSO-based optimization algorithm, where the warden only needs to know the values of its detection error probability.
	
	\subsubsection{Particle Swarm Optimization (PSO) Algorithm}
	In the PSO algorithm, $n$ particles travel in an $L$-dimensional search space to optimize the fitness function \cite{cheng2018a}. In our case, $L$ is equal to $1$. First, the particles are dispersed randomly in the search space, where each particle's position indicates a possible solution to the fitness function. Then, in every iteration, each particle evaluates the fitness and travels to a new place based on the history of the particle's best prior location and the global best position. The velocity $v_{i}$ and position $x_{i}$ of the $i_{\rm th}$ particle are updated using the following equations
	{\small \begin{subequations}
	\begin{align}
	&{v_{i}}\left( {k\! + \!1} \right)\! =\! \omega{v_{i}}\!\left( k \right) \!+\! {c_1}{R_1}\!\left( {{p_{i}}\!\left( k \right) \!-\! {x_{i}}\left( k \right)} \right) \!+\! {c_2}{R_2}\left( {{p_{g}}\!\left( k \right) \!-\! {x_{i}}\left( k \right)} \right), \label{slafdj1}\\
	&{x_{i}}\left( {k + 1} \right) = {x_{i}}\left( k \right) + {v_{i}}\left( {k + 1} \right), \label{slafdj2}
	\end{align}
	\end{subequations}}\noindent
	for $i = 1,2, \ldots ,m$, where $\omega $ is the inertia weight, $c_1$ and $c_2$ are acceleration
	constants, both $R_1$ and $R_2$ are uniformly distributed in $[0, 1]$, $ {p_i}$ is the best previous position, and $ {p_g}$ is the best global position in the swarm.
	\begin{algorithm}[t]
	\caption{The Particle swarm optimization for finding the optimal detection threshold for user $U_k$.} 
	\label{A2}
	\hspace*{0.02in} {\bf Input:}
	Input the size of swarm: $n$, maximum velocity: $v_{\rm max}$, inertia weight: $\omega$, acceleration constants: $c_1$ and $c_2$, iteration numbers: ${\rm M_{PSO}}$\\
	\hspace*{0.02in} {\bf Output:}
	The optimal detection threshold: ${\varepsilon^{opt} _{k}}$.
	\begin{algorithmic}[1]
	\State {\it Initialize the velocities and positions of $n$ particles, where the position means different ${\varepsilon _{k}}$ for user $k$}
	\For{Every particle in swarm}
	\State {\it Obtain $\xi_k$ with the help of \eqref{detection}}
	\EndFor
	\State Update personal and global best positions, and corresponding values
	\While{$k<{\rm M_{PSO}}$}
	\For{Every particle}
	\State {\it Obtain corresponding $\xi_k$ with the help of \eqref{detection}}
	\If{Current position is persional best position}
	\State Update its personal best position
	\EndIf
	\EndFor
	\State {\it Update global best positions (a set of detection thresholds for $K$ users), and corresponding fitness function values (detection error probability)}
	\For{Every particle}
	\State Update $v$ with the help of \eqref{slafdj1}
	\If{$v>v_{\rm max}$}
	\State $v \leftarrow v_{\rm max}$
	\EndIf
	\State Update $x$ with the help of \eqref{slafdj2}
	\If{{\it $x \le {\kappa ^2}$}}
	\State {\it $x={\kappa ^2}$}
	\EndIf
	\EndFor
	\EndWhile
	\State ${\varepsilon^{opt} _{k}}={\varepsilon_{k}}$
	\State \Return The global best position: ${\varepsilon^{opt} _{k}}$
	\end{algorithmic}
	\end{algorithm}
	
	To minimize ${\xi_{k}}$ by adjusting ${\varepsilon _{k}}$, the warden can initialize $n$ particles to search for the optimal detection threshold, ${\varepsilon^{opt} _{k}}$. The detailed PSO algorithm is given in Algorithm \ref{A2}. Because we have proven that ${\xi_{k}}$ is convex with respect to ${\varepsilon _{k}}$, Algorithm \ref{A2} can converge quickly. The total running time of Algorithm \ref{A2} can be expressed as $n{\rm M_{PSO}} T_{T}$, where $T_{T}$ is the time required by per particle in one iteration \cite{chraibi2020run}.
	
	\subsection{Game Theoretic Problem Formulation}
	{\color{black} Several multi-user communication scenarios have been studied with game-theoretic bargaining solutions, e.g., OFDMA channel allocation \cite{han2005fair}, bandwidth allocation for multimedia \cite{park2007bargaining}, and rate control for video coding \cite{wang2014generalized}. In cooperative games, players, i.e., $K$ users, aim to reach an agreement that is efficient and fair given the five axioms of PAR, IR, INV, IIA and SYM \cite{nash1950the}.} Each user $U_k$ has its own utility function and a minimum desired utility, called the disagreement point. The disagreement point is the minimum utility that each user expects. In our case, the resource is the transmitting and jamming power, and the utility function of user $U_k$ can be defined as $ {U_k}\left( {{P_{ak}},{P_{jk}}} \right) $. Let $ S = \left\{ {{U_1}, \ldots ,{U_K}} \right\} $ denote a joint utility set that is nonempty, convex, closed, and bounded, and $ D = \left\{ {U_1^{th}, \ldots ,U_K^{th}} \right\} $ denote the disagreement point set. The pair $ \left( {S,D} \right) $ defines the bargaining problem. 
	
	{\color{black} As an efficient resource allocation strategy, the NBS has been extensively deployed as a fair solution for the NBG \cite{nash1950the}.} With the help of NBS, our problem can be formulated as
	{\small \begin{subequations}
	\begin{align}
	\mathop {\max }\limits_{\{ {P_{jk}},{P_{ak}}\} }  \quad & \prod\limits_{k = 1}^K {\left( {{R_k} - R^{\rm th}_{k}} \right)}\\		
	{\rm s.t.}  \quad &\sum\limits_{k = 1}^K {{P_{jk}}}  \leqslant {P_J},\label{st1}\\
	&\sum\limits_{k = 1}^K {{P_{ak}}}  \leqslant {P_T},\label{st2}\\
	& \xi_k  \geqslant \xi^{\rm th}_k ,\forall k,\label{st3}
	\end{align}		
	\end{subequations}}\noindent
	where $R_k$ can be calculated as in \eqref{EQ-1}, $\xi^{\rm th}_k$ is a constant which is close to $1$, and $\xi_k$ can be obtained by \eqref{detection} under the optimal detection threshold.
	
	Note that our problem and solution are equivalent to standard Nash bargaining problem, and hence these five axioms that mentioned before are achievable intrinsically. However, the computational complexity required to find an NBS significantly increases as the number of users increases, which makes it difficult to obtain optimal allocation solutions given both $P_T$ and $P_J$ simultaneously. To solve the aforementioned issues, we propose the following two efficient algorithms and analyze their converge.
	\section{Joint Jamming and Transmitting Power Allocation Algorithm}\label{S5}
	In this section, we propose two algorithms, i.e., PPA and JTPA algorithms, for jammer-aided multi-antenna UAV covert communication networks. In the PPA algorithms, we let $n$ particles travel in a $2K$-dimensional convex search space. In the JTPA algorithm, we first optimize the jamming power allocation under the fixed transmit power allocation in Stage $1$. Then, in Stage $2$, we aim to determine the best transmit power allocation based on the results of Stage $1$. The overall algorithm stops until the convergence is achieved.
	\subsection{Converge Analysis}
	According to the definition of the covert rate as \eqref{aefjeal} in Appendix \ref{appendix2}, the cover rate increases monotonically with the increase of transmit power and decreases monotonically with the increase of jamming power. In the following, we first prove that the constraint set is a convex feasible set. Note that the power constraints, \eqref{st1} and \eqref{st2}, are convex with respect to $P_{jk}$ and $P_{ak}$, respectively. For \eqref{st3}, the Hessian matrix can be expressed as
	\begin{equation}
	{\nabla ^2}{\xi _k} = \left| {\begin{array}{*{20}{c}}
	{\frac{{{\partial ^2}{\xi _k}}}{{\partial {C_{1w}}^2}}}&{\frac{{{\partial ^2}{\xi _k}}}{{\partial {C_{1w}}\partial {C_{2w}}}}} \\ 
	{\frac{{{\partial ^2}{\xi _k}}}{{\partial {C_{2w}}\partial {C_{1w}}}}}&{\frac{{{\partial ^2}{\xi _k}}}{{\partial {C_{2w}}^2}}} 
	\end{array}} \right|.
	\end{equation}
	\begin{theorem}\label{appendixFF}
	The Hessian matrix of ${\xi _k}$ is semi-negative definite.
	\begin{IEEEproof}
	Please refer to Appendix \ref{appendixF}.
	\end{IEEEproof}
	\end{theorem}
	Therefore, we conclude that the constraint set is convex. However, from the expression of fitness function \eqref{EQ-1}, we can observe that the parameters $P_{jk}$ and $P_{ak}$ exist in one Fox's $H$-function. Thus, it is difficult, if not impossible, to derive the analytical expression of optimal $P_{jk}$ and $P_{ak}$. Fortunately, by regarding \eqref{detection} as the fitness function, many kinds of search algorithms, i.e., PSO, can be used to obtain the maximum $\xi_k$ and corresponding ${\varepsilon _{k}}$ because the search space is convex. Moreover, alternating optimization could also be used to trade off between the system performance and computational complexity.
	
	\subsection{PSO-based Power Allocation Algorithm}
	Because $\mathbb{P}_{FA}$ is not related to $P_{ak}$, and $\mathbb{P}_{MD}$ increases as $P_{ak}$ increases, \eqref{st3} limits the ceiling of $P_{ak}$. For given $P_{ak}$, ${\xi _k}$ is concave with the respect to $P_{jk}$ (see Appendix \ref{appendixF}). Thus, \eqref{st1}-\eqref{st3} actually limit the ranges of $P_{ak}$ and $P_{jk}$, which can be regarded as the boundary to PSO's search space. In the following, the PSO algorithm used for searching the maximum of objective function is proposed.
	\begin{figure}[t]
	\centering
	\includegraphics[scale=0.56]{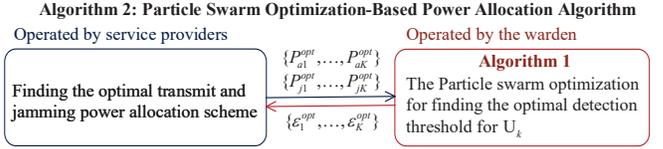}	
	\caption{The relationship of service providers' (i.e., UAV and jammer) PPA algorithm and the warden's detection threshold optimization algorithms: The transmit and jamming power will affect the optimal detection threshold, while the changing of detection threshold will change the optimal power allocation scheme simultaneously.}
	\label{A1G}
	\end{figure}
	
	To find the maximum $\prod\limits_{k = 1}^K {\left( {{R_k} - R^{\rm th}_{k}} \right)}$, we propose a PSO algorithm, denoted by Algorithm 2, under the particles' location limitation given in \eqref{st1}-\eqref{st3}\footnote{To save the space, we discuss the differences between Algorithm 2 and Algorithm 1 here, instead of giving the specific Algorithm 2.}. Different from Algorithm \ref{A2}, the output now is the optimal transmit and jamming power allocation, $\{P^{opt}_{j1}, \ldots, P^{opt}_{jK}\}$ and $\{P^{opt}_{a1}, \ldots, P^{opt}_{aK}\}$, the fitness function is $\prod\limits_{k = 1}^K {\left( {{R_k} - R^{\rm th}_{k}} \right)}$, and the dimension of searching space is $2K$ because the $2K$ coordinates of each particle represent one possible transmit and jamming power allocation scheme. Note that when the UAV and jammer are adjusting the joint power allocation schemes, the warden is optimizing its detection threshold based on the current transmit and jamming power. Furthermore, the detection threshold will also affect the Algorithm 2, as shown in Fig. \ref{A1G}. The total running time of Algorithm 2 is $T_{P}=2nKT_{T}{\rm M_{PSO}}$, where $T_{T}$ denotes the time required by per particle in one iteration, ${\rm M_{PSO}}$ is the iteration number, and $n$ denotes the size of swarm. 
	
	\subsection{Joint Two-Stage Power Allocation Algorithm}
	To reduce the computational complexity, we further propose an efficient two-stage power allocation algorithm based on the alternating optimization of the transmit and jamming power in an iterative manner. Note that maximizing $ \prod\limits_{k = 1}^K {\left( {{R_k} - R_k^{{\rm{th}}}} \right)}  $ equals to maximize $ \sum\limits_{k = 1}^K {\ln \left( {{R_k}-R_k^{{\rm{th}}}} \right)}  $.  Moreover, if $ {\left( {{R_k} - R_k^{{\rm{th}}}} \right)} $ is concave, $ {\ln \left( {{R_k} - R_k^{{\rm{th}}}} \right)} $ is also concave because ${\left( {\ln \left( {{F_k}} \right)} \right)''} = \frac{{{F_k}''{F_k} - {{\left( {{F_k}'} \right)}^2}}}{{{F_k}^2}}$, where $ {F_k} \buildrel \Delta \over = {R_k} - R_k^{{\rm{th}}} $, $ {\left(  \cdot  \right){'}} $ denotes the first order derivative and $ {\left(  \cdot  \right){''}} $ denotes the second order derivative.
	
	\subsubsection{Stage 1: Optimization of Jamming Power}
	For any given transmit power allocation, the jamming power allocation problem can be formulated as
	{\small \begin{subequations}\label{prob1}
	\begin{align}
	\mathop {\max }\limits_{\{ {P_{jk}}\} }  \quad & \sum\limits_{k = 1}^K {\ln \left( {{R_k} - R_k^{{\rm{th}}}} \right)} \label{aegfe2sa}\\	
	{\rm s.t.}  \quad &\eqref{st1}, \eqref{st2}, \eqref{st3}.
	\end{align}	
	\end{subequations}}\noindent
	If the transmit power allocated to a user by the UAV is too high, the warden's detection error probability for that user will decrease, resulting in a failure of covert communication. Therefore, the jammer needs to allocate the appropriate amount of power to interfere the detection of the warden. Although the jamming signals will also affect the user's SINR and cause a decrease in the covert rate, the interference that it causes to the warden allows the UAV to utilize a higher transmit power for data transmission, thus ultimately increasing the cover rate after transmit power optimization compared to the absence of jamming signals \cite{huang2021jamming}. Note that \eqref{aegfe2sa} can be solved using swarm intelligence algorithms \cite{derrac2011a} or the method of Lagrange multipliers \cite{rockafellar1993lagrange}, which have been widely studied \cite{liu2021a,yu2020power,zhang2019rate,chien2020joint}.
	
	\subsubsection{Stage 2: Optimization of Transmit Power}
	In Stage 2, we design the optimal transmit power allocation scheme under the fixed jamming power. The problem can be formulated as
	{\small \begin{subequations}\label{prob2}
	\begin{align}
	\mathop {\max }\limits_{\{ {P_{ak}}\} }  \quad & \sum\limits_{k = 1}^K {\ln \left( {{R_k} - R_k^{{\rm{th}}}} \right)} \\	
	{\rm s.t.}  \quad &\eqref{st1}, \eqref{st2}, \eqref{st3}.
	\end{align}		
	\end{subequations}}\noindent
	This problem can be solved by the similar method as in Stage 1.	Although an increase in transmit power may decrease the detection error probability of warden, the jamming power allocation scheme proposed in Stage 1 can help to increase the detection error probability.
	
	\subsubsection{Joint Two-Stage Power Allocation Algorithm}
	Based on the results presented in the previous two subsections, we propose a two-stage alternating iterative algorithm. Specifically, let ${\bf P_{j}}=\{P_{j1}, \ldots, P_{jK}\}$ and ${\bf P_{a}}=\{P_{a1}, \ldots, P_{aK}\}$. In the ${(r+1)}$ iteration, we use a fixed ${\bf P}^{(r)}_{a}$ to obtain ${\bf P}^{(r+1)}_{j}$ in Stage 1. Then, with the help of ${\bf P}^{(r+1)}_{j}$, ${\bf P}^{(r+1)}_{a}$ is solved in Stage 2. The overall algorithm stops iterating when the difference between the objective function values obtained in two iterations is less than a threshold. The procedure of the JTPA is summarized in Algorithm \ref{A5}. 
	%Let $M_{A}$ denote the maximum number of iterations that allows JTPA to converge.	
	\setcounter{algorithm}{2}
	\begin{algorithm}[t]
	\caption{Joint two-stage transmit and jamming power allocation algorithm.} 
	\label{A5}
	\hspace*{0.02in} {\bf Input:}
	Input the size of swarm: $n$, dimension of searching space: $K$, maximum velocity: $v_{\rm max}$, inertia weight: $\omega$, acceleration constants: $c_1$ and $c_2$, iteration numbers: ${\rm M_{PSO}}$\\
	\hspace*{0.02in} {\bf Output:}
	The transmit and jamming power allocation schemes: {\small $\{P^{opt}_{j1}, \ldots, P^{opt}_{jK}\}$} and {\small $\{P^{opt}_{a1}, \ldots, P^{opt}_{aK}\}$}
	\begin{algorithmic}[1]
	\State Initialize {\small ${\bf P}^{(0)}_{a}$} and {\small ${\bf P}^{(0)}_{j}$}, calculate the objective function as {\small $\sum\limits_{k = 1}^K {\ln \left( {{R_k} - R_k^{{\rm{th}}}} \right)}^{(0)}$}, set the iteration number: {\small $r=0$}, set the threshold: {\small $\rho$}.
	\Repeat 
	\State Stage 1: For fixed {\small ${\bf P}^{(r)}_{a}$}, solve \eqref{prob1} to obtain {\small ${\bf P}^{(r+1)}_{j}$}.
	\State Stage 2: For fixed {\small ${\bf P}^{(r+1)}_{j}$}, solve \eqref{prob2} to obtain {\small ${\bf P}^{(r+1)}_{a}$}.
	\State Denote the objective function as {\small $\sum\limits_{k = 1}^K \!\!{\ln\!\left( {{R_k} \!-\! R_k^{{\rm{th}}}} \right)}^{(r+1)}$}.
	\State {\small $r=r+1$}.
	\Until {\small $\left| {\sum\limits_{k = 1}^K {\ln \left( {{R_k} - R_k^{{\rm{th}}}} \right)}^{(r)}-\sum\limits_{k = 1}^K {\ln \left( {{R_k} - R_k^{{\rm{th}}}} \right)}^{(r-1)} } \right|<\rho$}
	\State {\small $\{P^{opt}_{j1}, \ldots, P^{opt}_{jK}\}=\{P_{j1}, \ldots, P_{jK}\}$} and {\small $\{P^{opt}_{a1}, \ldots, P^{opt}_{aK}\}=\{P_{a1}, \ldots, P_{aK}\}$}.
	\State \Return The global best position: {\small $\{P^{opt}_{j1}, \ldots, P^{opt}_{jK}\}$} and {\small $\{P^{opt}_{a1}, \ldots, P^{opt}_{aK}\}$}.
	\end{algorithmic}
	\end{algorithm}
	
\begin{figure}[t]
	\centering
	\includegraphics[scale=0.6]{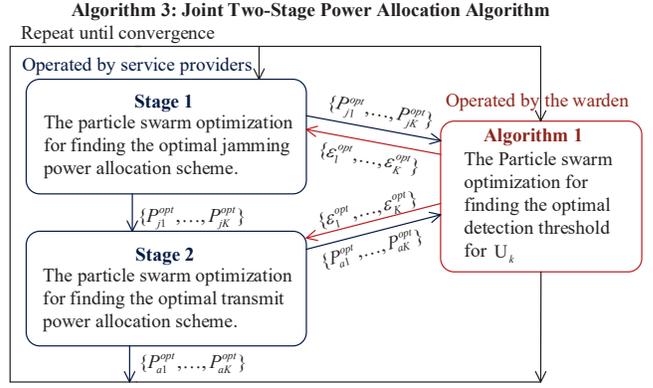}	
	\caption{The relationship of service providers' (i.e., UAV and jammer) JTPA algorithm and the warden's detection threshold optimization algorithms: In Stage 1, the service providers adjust the jamming power, and the warden will obtain a set of corresponding detection thresholds. In Stage 2, the service provider will obtain the optimal transmit power based on the current thresholds. Then we go back to Stage 1 again until the convergence is achieved.}
	\label{Algorithms}
	\end{figure}

	While the UAV and jammer are using Algorithm \ref{A5} for power allocation, the warden is also using Algorithm \ref{A2} for optimal detection threshold design. Since the UAV, jammer, and the warden are working simultaneously and interacting with each other, we give Fig. \ref{Algorithms} to show the relationship of the five algorithms proposed in this paper.
	
	\section{Numerical Results}\label{S6}
	In this section, simulation results are presented to verify the proposed PPA and JTPA algorithms for jammer-aided multi-antenna UAV covert communication networks. We set $K=3$, and the locations of the three users, the jammer, the UAV, and the warden are ${\bf q}_1=[0,10,0]^T$, ${\bf q}_2=[7,14,0]^T$, ${\bf q}_3=[5,20,0]^T$, ${\bf q}_j=[7,24,0]^T$, ${\bf q}_a=[5,13,20]^T$, and ${\bf q}_w=[5,16,0]^T$, respectively. The coefficients of path loss model are $\alpha_{a1}=1$, $\alpha_{a2}=1.2$, $\alpha_{a3}=1.3$, $\alpha_{aw}=1.2$, $\alpha_{j1}=1.7$, $\alpha_{j2}=1.6$, $\alpha_{j3}=1.8$, $\alpha_{jw}=1.7$. The AWGN power is $ {{\kappa ^2}}=3 $ ${\rm dB}$. For parameters of Fisher-Snedecor $\mathcal{F}$ fading model, we set $m_{f1}=2$, $m_{f2}=3$, $m_{f3}=5$, $m_{fw}=3$, $m_{s1}=4$, $m_{s2}=4$, $m_{s3}=5$, $m_{sw}=4$, ${\bar z}_{1}=-10$ ${\rm dB}$, ${\bar z}_{2}=-12$ ${\rm dB}$, ${\bar z}_{3}=-13$ ${\rm dB}$, and ${\bar z}_{w}=-11$ ${\rm dB}$. For parameters of FTR fading model, we set $m_{1}=4$, $m_{2}=3$, $m_{3}=5$, $m_{w}=4$, $K_{1}=4$, $K_{2}=5$, $K_{3}=5$, $K_{w}=3$, $\Delta_{1}=0.5$, $\Delta_{2}=0.5$, $\Delta_{3}=0.4$, $\Delta_{w}=0.4$, $2\sigma^2_1(1+K_1)=-10$ ${\rm dB}$, $2\sigma^2_2(1+K_2)=-15$ ${\rm dB}$, $2\sigma^2_3(1+K_3)=-11$ ${\rm dB}$, and $2\sigma^2_w(1+K_w)=-10$ ${\rm dB}$. 
	%	\begin{figure}[t]
	%		\begin{minipage}[t]{0.47\linewidth}
	%			\centering
	%			\includegraphics[width=1.1\textwidth]{DEP.eps}	
	%			\caption{The detection error probability, $\xi_k$ $(k=1,2,3)$, versus the detection threshold with or without our proposed JTPA algorithm, with $P_T=20$, $P_J=20$, and $\xi^{\rm th}_k=95\%$.}
	%			\label{DEP}
	%		\end{minipage}
	%		\hfill
	%		\begin{minipage}[t]{0.47\linewidth}
	%			\centering
	%			\includegraphics[width=1.1\textwidth]{CRRTH.eps}	
	%	\caption{The NBS's utility function versus with the jammer's available total power under the different detection error thresholds, with $P_T=30$ ${\rm dB}$.}
	%\label{CRRTH}
	%		\end{minipage}
	%	\end{figure}
	\begin{figure}[t]
	\centering
	\includegraphics[scale=0.5]{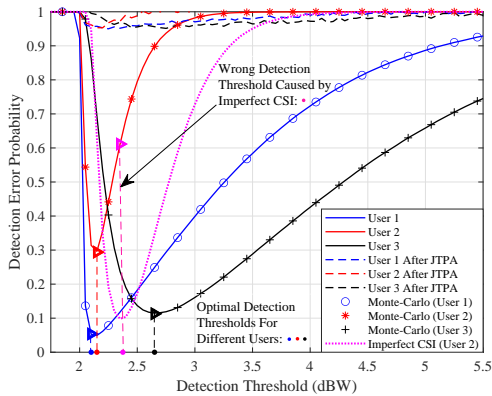}	
	\caption{{\color{black} The detection error probability, $\xi_k$ $(k=1,2,3)$, versus the detection threshold with or without our proposed JTPA algorithm, with $P_T=20$ ${\rm dBW}$, $P_J=20$ ${\rm dBW}$, and $\xi^{\rm th}_k=95\%$.}}
	\label{DEP}
	\end{figure}
	
	Figure \ref{DEP} shows the detection error probability, $\xi_k$ $(k=1,2,3)$, versus the detection threshold with or without our proposed JTPA algorithm, with $P_T=20$ ${\rm dBW}$, $P_J=20$ ${\rm dBW}$, and $\xi^{\rm th}_k=95\%$. Under the fixed total power parameters (i.e., $P_T$ and $P_J$), we can observe that each curve has its optimal detection threshold which can be captured by the warden. With the help of Algorithm \ref{A2}, the optimal detection threshold for each user obtained by the warden is marked with stars. Note that each link's corresponding detection error probability fails to achieve the covertness requirements. Thus, it is necessary to use our PPA or JTPA algorithms to ensure the covertness of communication. As shown by the dotted lines in Fig. \ref{DEP}, all users' communication can make the warden's detection error probability reach more than 95\% and be turned into secure ones, even under the warden's optimal detection threshold. Specifically, the error probabilities of users $U_k$ $(k=1,2,3)$ increase $1800\%$, $196\%$ and $691\%$, respectively. These results demonstrate the effectiveness of our proposed JTPA algorithm. {\color{black}Furthermore, if the CSI is imperfectly estimated, the warden will choose the wrong detection threshold and fail to achieve the minimal DEP. For example, when ${e_{jw}}={e_{aw}}=-10$ ${\rm dB}$, the corresponding DEP is $107.9\%$ higher than the minimal one.}
	\begin{figure}[t]
	\centering
	\includegraphics[scale=0.5]{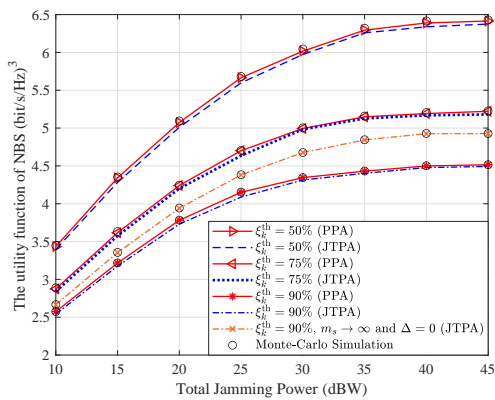}	
	\caption{The NBS's utility function versus with the jammer's available total power under the different detection error thresholds, with $P_T=30$ ${\rm dBW}$.}
	\label{CRRTH}
	\end{figure}
	
	Figure \ref{CRRTH} depicts the NBS's utility function versus the jammer's available total power under the different detection error thresholds, with $P_T=20$ ${\rm dBW}$. More specifically, the utility function increases with the increase of the available total power of the jammer, e.g., we can observe that the utility function increases by $78\%$ when $P_{J}$ increases from $10$ ${\rm dBW}$ to $30$ ${\rm dBW}$ with $\xi^{\rm th}_k=75\%$. 	
	 Another interesting insight is that the utility is higher when the $\xi^{\rm th}_k$ is lower. For example, when $\xi^{\rm th}_k$ decreases from $90\%$ to $50\%$ with $P_{J}=30$ ${\rm dBW}$, the utility function increases by $39\%$. The reason is that a high $\xi^{\rm th}_k$ means that the system has high standards for covertness. Thus, the UAV has to allocate lower transmit power to users to meet the covertness requirements, especially when the available total power of the jammer is scarce or the jammer is located in a relatively distant position. 
	 In contrast, we can deploy a higher UAV's transmit power $P_T$ to enhance the system's utility when the jammer has sufficient power to dramatically degrade the warden's detection capability and provide a greater margin for the system's utility. {\color{black}For special small-scale fading cases, i.e., $m_s \to \infty$ (Fisher-Snedecor $\mathcal{F}$ $\to$ Nakagami-$m$) and $\Delta = 0$ (FTR fading $\to$ Rician shadowed fading), the total utility is higher because of better channel quality.}
	 However, the utility no longer increases with the jammer's total power in the high-jamming power range, because the total jamming power at this range is sufficient for three users. Furthermore, we can observe that the utility achieved by PPA algorithm is sightly larger then that achieved by JTPA algorithm. The reason is that there is no iteration in the PPA algorithm. However, JTPA is more efficient because PPA algorithm requires 2$K$ particles in the search space simultaneously. 
	%	\begin{figure}[t]
	%		\begin{minipage}[t]{0.47\linewidth}
	%			\centering
	%			\includegraphics[width=1.1\textwidth]{CRP.eps}	
	%			\caption{The NBS's utility function versus the total transmit power of UAV, under different sets of total transmit powers of UAV, with $\xi^{\rm th}_k = 90\%$.}
	%			\label{UAVL}
	%		\end{minipage}
	%		\hfill
	%		\begin{minipage}[t]{0.47\linewidth}
	%			\centering
	%			\includegraphics[width=1.1\textwidth]{UAVL.eps}	
	%			\caption{The NBS's utility versus the UAV's location, with $P_T=20$ ${\rm dB}$, $P_J=20$ ${\rm dB}$, ${\bf q}_1=[3,15,0]^T$, ${\bf q}_2=[6,18,0]^T$, ${\bf q}_3=[8,12,0]^T$, and $\xi^{\rm th}_k = 90\%$.}
	%			\label{CRP}
	%		\end{minipage}
	%	\end{figure}
	\begin{figure}[t]
	\centering
	\includegraphics[scale=0.5]{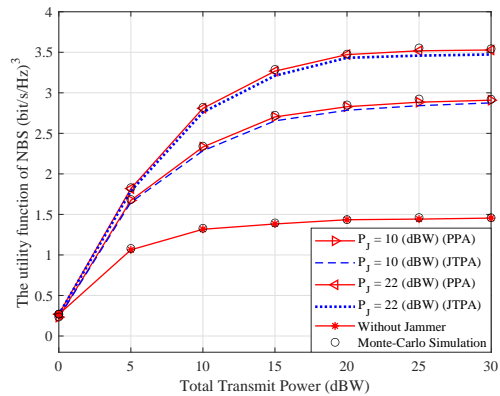}	
	\caption{The NBS's utility function versus the total transmit power of UAV, under different sets of total transmit powers of UAV, with $\xi^{\rm th}_k = 90\%$.}
	\label{UAVL}
	\end{figure}
	
	As a supplement, Fig. \ref{UAVL} shows the NBS's utility function versus the total transmit power of UAV, under different sets of total transmit powers of UAV, with $\xi^{\rm th}_k = 90\%$. Figure \ref{UAVL} illustrates that a higher total transmit power $P_T$ can achieve a larger utility function of NBS. It is interesting that when the total transmit power is small, the utility functions at different total jamming powers are very close to each other. The reason is that the jamming power required to achieve a high detection error probability is low when the transmit power is low. When the total transmit power is larger, we can observe that the difference between utility functions under different total jamming power is larger. For example, when $P_{T}=30$ ${\rm dBW}$, the utility function increases $22\%$ when $P_J$ increases from $10$ ${\rm dBW}$ to $20$ ${\rm dBW}$. Furthermore, due to the covertness requirements, we can observe that there is an upper bound of the UAV's transmit power where the utility no longer increases with its total power. The case without jammer is also shown in Fig. \ref{UAVL}. We can observe that the utility achieved with the help of jammer is significantly larger. Combined with the result in Fig. \ref{CRRTH}, it is proved that the jammer's total power and the UAV's transmit power jointly determine the system's utility, and our proposed PPA and JTPA algorithms are effective.
	
	\begin{figure}[t]
	\centering
	\includegraphics[scale=0.5]{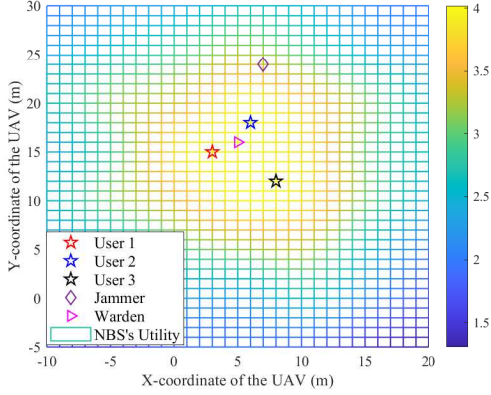}	
	\caption{{\color{black} The NBS's utility versus the UAV's location, with $P_T=20$ ${\rm dBW}$, $P_J=20$ ${\rm dBW}$, ${\bf q}_1=[3,15,0]^T$, ${\bf q}_2=[6,18,0]^T$, ${\bf q}_3=[8,12,0]^T$, and $\xi^{\rm th}_k = 90\%$.}}
	\label{CRP}
	\end{figure}
	To further illustrate the impact of the UAV's location on the NBS's utility, Fig. \ref{CRP} plots the relationship between the horizontal plane coordinate (the height of UAV is fixed) and the NBS's utility function, with $P_T=20$ ${\rm dBW}$, $P_J=20$ ${\rm dBW}$, ${\bf q}_1=[3,15,0]^T$, ${\bf q}_2=[6,18,0]^T$, ${\bf q}_3=[8,12,0]^T$ and $\xi^{\rm th}_k = 90\%$. {\color{black} For each point shown in Fig. \ref{CRP}, the PPA and JTPA algorithms are used to maximize the covert rate, which also proves the effectiveness of our algorithms even when the UAV's position varies over a wide range. Furthermore, in Fig. \ref{CRP}, we can observe that the value of the utility function can be further improved by UAV's trajectory planning optimization. Within the flying range shown in Fig. \ref{CRP}, The utility function when the UAV is in the optimal position is $167\%$ higher than that when the UAV is in the worst position. The detailed analysis is left for the future work.}
	%=============================================================
	\section{Conclusion}\label{S7}
	A jammer-aided multi-antenna UAV covert communication system is investigated. We used the Fisher-Snedecor ${\mathcal F}$ fading and FTR fading to derive the exact PDF and CDF of the SINR. Important covert performance metrics including detection error probability and covert rate were derived. With the help of the obtained performance metrics, we formulated the joint transmit and jamming power allocation problem as a NBG to maximize the user's covert rate with limited transmit and jamming power and ensure the covertness of communication simultaneously. To solve the formulated problem, we proposed PPA and JTPA algorithms, under the warden's optimal detection threshold. PSO were used by the warden in finding the optimal detection threshold. Furthermore, all formulated problems are proven to be convex, and the running times were investigated. Numerical results illustrated that the jammer's total power and the UAV's transmit power can be jointly allocated to improve the covert rate, and our proposed algorithms are effective. For future works, we can extend the power allocation algorithms to joint power and the three-dimension trajectory optimization algorithms, because the UAV may fly away from the warden to weaken the detection channels.
	
	\begin{appendices}
	\section{Proof of Lemma \ref{Lemma-1}}\label{appendix1} 
	\renewcommand{\theequation}{A-\arabic{equation}}
	\setcounter{equation}{0}
	\subsubsection{Proof of PDF}
	Let $U_k \triangleq \frac{\gamma_k }{{{C_{1k}}}} = \frac{{{X_k}}}{T_k}$, where $T_k \triangleq {\kappa ^2} + {C_{2k}}{Z_k}$.
	The PDF of $T_k$ can be expressed as ${f_{T_k}}\!\left( t \right) =\frac{1}{{{C_{2k}}}}{f_{Z_k}}\!\left( {\frac{{t - {\kappa ^2}}}{{{C_{2k}}}}} \right)$.
	With the help of \eqref{PDFFTR}, we can obtain the PDF of $U_k$ as
	{\small  \begin{align}\label{EQ-AP-A-1}
	&{f_{U_k}}\!\left( u \right) =\! \int_0^\infty \! \!\!{x{f_{X_k}}\left( {ux} \right){f_{T_k}}\!\left( x \right){\rm d}x}\notag\\
	&= \!\frac{{{m_k}^{{m_k}}}}{{\Gamma\!\left( {{m_k}} \right)}}\sum\limits_{j = 0}^M {\frac{{{K_k}^j{\alpha _{kj}}{u^j}{m_{fk}}^{{m_{fk}}}{{\left( {{m_{sk}} - 1} \right)}^{{m_{sk}}}}{{\bar z}_k}^{{m_{sk}}}}I_1}{{j!\Gamma\!\left( {j \!+\! 1} \right){{\left( {2\sigma _k^2} \right)}^{j + 1}}{C_{2k}}^{{m_{fk}}}B\!\left( {{m_{fk}},{m_{sk}}} \right)}}}  ,
	\end{align}}\noindent
	where 
	{\small \begin{equation}
	{I_1}\! =\!\! \int_0^\infty \!\!\!\! {\frac{{{x^{j + 1}}{{\left( {x - {\kappa ^2}} \right)}^{{m_{fk}} - 1}}\exp \left( { - ux{{\left( {2\sigma _k^2} \right)}^{ - 1}}} \right)}{\rm d}x}{{{{\left(\! {{m_{fk}}x{C_{2k}^{ - 1}}\! -\! {m_{fk}}{\kappa ^2}{C_{2k}^{ - 1}} \!+\! \left( {{m_{sk}} \!-\! 1} \right){{\bar z}_k}} \!\right)}^{{m_{fk}} + {m_{sk}}}}}}}.
	\end{equation} }\noindent
	With the help of \cite[eq. (01.03.07.0001.01)]{web}, we can re-write $I_1$ as
	{\small \begin{align}\label{I1newpdf}
	&	{I_1} = \frac{1}{{2\pi i}}\int_{{{\cal L}_1}} {\Gamma \left( { - {s_1}} \right)} {\left( {\frac{u}{{2\sigma _k^2}}} \right)^{{s_1}}}\notag\\
	&\times \!\int_0^\infty \!\!\!{\frac{{{x^{1 + {s_1} + j}}{{\left( {x - {\kappa ^2}} \right)}^{{m_{fk}} - 1}}{\rm d}x{\rm d}{s_1}}}{{{{\left( {{m_{fk}}x{C_{2k}^{ - 1}}\! - \!{m_{fk}}{\kappa ^2}{C_{2k}^{ - 1}}\! +\! \left( {{m_{sk}} \!-\! 1} \right){{\bar z}_k}} \right)}^{{m_{fk}} + {m_{sk}}}}}}},
	\end{align}}\noindent
	where the integration path of $\mathcal{L}_1$ goes from $\sigma_L -i\infty $ to $\sigma_L+i\infty $ and $\sigma  \in \mathbb{R}$. With the help of \cite[eq. (3.197.1)]{gradshteyn2007}, \cite[eq. (9.113)]{gradshteyn2007} and \cite[eq. (8.384.1)]{gradshteyn2007}, the integration part in \eqref{I1newpdf} can be solved. Substituting $I_1$ into \eqref{EQ-AP-A-1}, after some algebraic manipulations, we further express ${f_U}\left( u \right)$ as \eqref{fubott}, shown at the bottom of the next page.
	%----------------------------------------------------------------------------------------------------------------------------
	%----------------------------------------------------------------------------------------------------------------------------
	\newcounter{mycount4}
	\begin{figure*}[b]
	\normalsize
	\setcounter{mycount4}{\value{equation}}
	\hrulefill
	\vspace*{4pt}
	{\small {\color{black} \begin{align}\label{fubott}
	{f_U}\left( u \right) =& \frac{{{m_k}^{{m_k}}}}{{u\Gamma \left( {{m_k}} \right)\Gamma \left( {{m_{fk}}} \right)\Gamma \left( {{m_{sk}}} \right)}}{\left( {\frac{{{m_{fk}}{\kappa ^2}}}{{{\Omega _k}}}} \right)^{{m_{fk}}}}\sum\limits_{j = 0}^\infty  {\frac{{{K_k}^j{\alpha _{kj}}}}{{\Gamma \left( {j + 1} \right)j!}}} {\left( {\frac{1}{{2\pi i}}} \right)^2}\int_{{{\cal L}_1}} {\int_{{{\cal L}_2}} {{{\left( {\frac{{ {m_{fk}}{\kappa ^2}- {\Omega _k}}}{{{\Omega _k}}}} \right)}^{{s_2}}}} } 
	\notag\\&\times\!\!
	\frac{{\Gamma \left( {1 + j + {s_1}} \right)\Gamma \left( {{m_{sk}} + {s_1}} \right)\Gamma \left( {{m_{fk}} + {s_2}} \right)\Gamma \left( {{m_{sk}} + {m_{fk}} + {s_2}} \right)\Gamma \left( { - {s_2}} \right)}}{{\Gamma \left( {{m_{fk}} + {m_{sk}} + {s_1} + {s_2}} \right)}}{\left( {\frac{{2\sigma _k^2}}{{{\kappa ^2}u}}} \right)^{{s_1}}}{\rm d}{s_2}{\rm d}{s_1}
	\end{align}}}\noindent	
	\setcounter{equation}{\value{mycount4}}
	\end{figure*}
	\addtocounter{equation}{1}
	%----------------------------------------------------------------------------------------------------------------------------
	%----------------------------------------------------------------------------------------------------------------------------
	
	Then, with the help of \eqref{fubott} and \cite[eq. (A-1)]{mathai2009h}, ${f_U}\left( u \right)$ can be expressed as 
	{\small {\color{black} \begin{equation}
	\begin{split}
	&{f_U}\!\left( u \right)\!
	= \frac{{{m_k}^{{m_k}}}}{{u\Gamma\!\left( {{m_k}} \right)\Gamma\!\left( {{m_{fk}}} \right)\Gamma\!\left( {{m_{sk}}} \right)}}{\left(\! {\frac{{{m_{fk}}{\kappa ^2}}}{{{\Omega _k}}}} \!\right)^{{m_{fk}}}}\sum\limits_{j = 0}^\infty  {\frac{{{K_k}^j{\alpha _{kj}}}}{{\Gamma \left( {j \!+\! 1} \right)j!}}} \\
	&\times\!\! H_{1,0:0,2:1,2}^{0,0:2,0:2,1}\!\!\!\left(\!\!\!\! \!{\left. {\begin{array}{*{20}{c}}
	{\frac{{{\kappa ^2}u}}{{2\sigma _k^2}}}\\
	{\frac{{{\Omega _k}}}{{{m_{fk}}{\kappa ^2} - {\Omega _k}}}}
	\end{array}}\!\!\!\!\right|\!\!\!\!\!\begin{array}{*{20}{c}}
	{\left( {{m_{fk}} \!+ \!{m_{sk}};\!1,\!1} \right)\!\!:-:\!\!\left( {1,\!1} \right)}\\
	{ - \!\!: \!\!\left(\! {1\! + \!j,\!1} \!\right)\!\left(\! {{m_{sk}},\!1} \!\right)\!\! :\!\!\left( \!{{m_{fk}},\!1} \!\right)\!\left(\! {{m_{sk}} \!+\! {m_{fk}},\!1}\! \right)}
	\end{array}} \!\!\!\!\!\right)\!\!.
	\end{split}
	\end{equation}}}\noindent
	Hence, since ${f_\gamma }\left( \gamma  \right) = \frac{1}{{{C_{1k}}}}{f_U}\left( {\frac{\gamma }{{{C_{1k}}}}} \right)$, we can derive \eqref{EQ-PDF}.
	\subsubsection{Proof of CDF}
	With the help of the definition of CDF, we have ${F_\gamma }\!\left( \gamma  \right) 	= \int_0^\gamma  \!{{f_\gamma }\left( x \right){\rm d}x}$. Substituting \eqref{EQ-PDF} into the definition, the CDF of $\gamma$ can be re-written as \eqref{aefaef23}, shown at the bottom of the next page, where $I_{A2}$ can be easily solved as $I_{A2}=\int_0^\gamma  {{x^{{s_1} - 1}}{\rm d}x} = \frac{1}{{{s_1}}}{\gamma ^{{s_1}}} = \frac{{\Gamma \left( {{s_1}} \right)}}{{\Gamma \left( {{s_1} + 1} \right)}}{\gamma ^{{s_1}}}$. Substituting $I_{A2}$ into \eqref{aefaef23} and using \cite[eq. (A-1)]{mathai2009h}, we can derive \eqref{EQ-CDF}. The proof is completed.
	
	%----------------------------------------------------------------------------------------------------------------------------
	%----------------------------------------------------------------------------------------------------------------------------
	\newcounter{mycount5}
	\begin{figure*}[b]
	\normalsize
	\setcounter{mycount5}{\value{equation}}
	\hrulefill
	\vspace*{4pt}
	{\small {\color{black} \begin{align}\label{aefaef23}
	{F_\gamma }\!\left( \gamma  \right)&\frac{{{m_k}^{{m_k}}}}{{\Gamma \left( {{m_k}} \right)\Gamma \left( {{m_{fk}}} \right)\Gamma \left( {{m_{sk}}} \right)}}{\left( {\frac{{{m_{fk}}{\kappa ^2}}}{{{\Omega _k}}}} \right)^{{m_{fk}}}}\sum\limits_{j = 0}^\infty  {\frac{{{K_k}^j{\alpha _{kj}}}}{{\Gamma \left( {j + 1} \right)j!}}} {\left( {\frac{1}{{2\pi i}}} \right)^2}\int_{{{\cal L}_1}} {\int_{{{\cal L}_2}} {{I_{{A_2}}}{{\left( {\frac{{{\Omega _k}}}{{{m_{fk}}{\kappa ^2} - {\Omega _k}}}} \right)}^{{s_2}}}} } 
	\notag\\&\times\!\!
	\frac{{\Gamma \left( {1 + j - {s_1}} \right)\Gamma \left( {{m_{sk}} - {s_1}} \right)\Gamma \left( {{m_{fk}} - {s_2}} \right)\Gamma \left( {{m_{sk}} + {m_{fk}} - {s_2}} \right)\Gamma \left( {{s_2}} \right)}}{{\Gamma \left( {{m_{fk}} + {m_{sk}} - {s_1} - {s_2}} \right)}}{\left( {\frac{{{\kappa ^2}}}{{2\sigma _k^2}}\frac{1}{{{C_{1k}}}}} \right)^{{s_1}}}{\rm d}{s_2}{\rm d}{s_1}
	\end{align}}}\noindent
	\setcounter{equation}{\value{mycount5}}
	\end{figure*}
	\addtocounter{equation}{1}
	%----------------------------------------------------------------------------------------------------------------------------
	%----------------------------------------------------------------------------------------------------------------------------
	{\color{black} \section{Proof of Propositions \ref{ApproProp1}-\ref{ApproProp3}}\label{NewAppen} 
		\renewcommand{\theequation}{B-\arabic{equation}}
		\setcounter{equation}{0}
		\subsection{Proof of Propositions \ref{ApproProp1}}\label{NewAppen1}
		With the help of \cite[eq. (A-1)]{mathai2009h}, the Multivariate Fox's $H$-function in \eqref{EQ-CDF} can be expressed in the Mellin-Barnes integral form. When the interference from the jamming signal is low, i.e., $ {\Omega _k} \to 0$, we have $ {\frac{{{\Omega _k}}}{{{m_{fk}}{\kappa ^2} - {\Omega _k}}}} \to 0$. Thus, the Mellin-Barnes integrals can be approximated by evaluating the residue at the minimum pole on the right-hand side \cite[Theorem 1.11]{mathai2009h}, and we have
		{\small \begin{align}\label{ageagh1}
				&	F_{{\gamma _k}}^{LJ}\left( \gamma  \right) = \frac{{{m_k}^{{m_k}}}}{{\Gamma \left( {{m_k}} \right)}}{\left( {\frac{{{m_{fk}}{\kappa ^2}}}{{{m_{fk}}{\kappa ^2} - {\Omega _k}}}} \right)^{{m_{fk}}}}\sum\limits_{j = 0}^\infty  {\frac{{{K_k}^j{\alpha _{kj}}}}{{\Gamma \left( {j + 1} \right)j!}}} 
				\notag\\&\times
				\frac{1}{{2\pi i}}\int_{{{\cal L}_1}} {\frac{{\Gamma \left( {1 + j + {s_1}} \right)\Gamma \left( { - {s_1}} \right)}}{{\Gamma \left( {1 - {s_1}} \right)}}} {\left( {\frac{{2\sigma _k^2{C_{1k}}}}{{{\kappa ^2}{\gamma}}}} \right)^{{s_1}}}d{s_1}.
		\end{align}}\noindent
		Using \cite[eq. (9.210.1)]{gradshteyn2007}, we obtain \eqref{afhlke} to complete the proof.
		\subsection{Proof of Propositions \ref{ApproProp2}}\label{NewAppen2}
		When the transmit power is high, i.e., $ {{C_{1k}}}\to \infty$, we have $ {\frac{{{\kappa ^2}\gamma }}{{2\sigma _k^2{C_{1k}}}}} \to 0$. Following the similar methods used in Appendix \ref{NewAppen1}, we can approximate the Mellin-Barnes integrals by evaluating the residue at the minimum pole on the right-hand side. However, $ \min \left\{ {1 + j,{m_{sk}}} \right\} $ is different because $j$ goes from $0$ to $M$. Thus, when $m_{sk}$ is not an integer, we can find $j_s$ which satisfies $ {j_s} + 1 < {m_{sk}} < {j_s} + 2 $. When $ j \in \left[ {0,{j_s}} \right] $, the minimum pole is $1+j$. When $ j \in \left[ {{j_s} + 1,M} \right] $, the minimum pole is $m_{sk}$. Note that when $m_{sk}$ is an integer, we can set $ {j_s} \triangleq {m_{sk}} - 1 $. In this case, we only need to calculate the double residue at ${m_{sk}}$. Then, we can obtain \eqref{safg123}, which completes the proof. 
		
		\subsection{Proof of Propositions \ref{ApproProp3}}\label{NewAppen3}
		With the help of \eqref{ageagh1}, when the transmit power is high, i.e., $ {\frac{{{\kappa ^2}\gamma }}{{2\sigma _k^2{C_{1k}}}}} \to 0$, the Mellin-Barnes integral over $\mathcal{L}_1$ can be approximated by the similar methods in Appendix \ref{NewAppen1}. Thus, we can derive \eqref{1dsad} to complete the proof.}
	
	\section{Proof of Theorem \ref{IEEEPROOF3}}\label{appendix3} 
	\renewcommand{\theequation}{C-\arabic{equation}}
	\setcounter{equation}{0}
	For the term of $\mathbb{P}_{AF}$, we have $\mathbb{P}_{AF} = \Pr \left( {{\kappa ^2} + {C_{2w}}Z_w > {\varepsilon _{k}}} \right)$. Denote that $T_w \triangleq {\kappa ^2} + {C_{2w}} Z_w$, and we can derive the CDF of $T_w$ as
	{\small \begin{align}\label{CDFTW}
	{F_{T_w}}\left( t \right) &= \Pr \left( {{T_w} < t} \right) = \Pr \left( {{\kappa ^2} + {C_{2w}}Z_w < t} \right)\notag\\&
	= \Pr \left( {Z_w < \frac{{t - {\kappa ^2}}}{{{C_{2w}}}}} \right) = {F_{Z_w}}\left( {\frac{{t - {\kappa ^2}}}{{{C_{2w}}}}} \right),
	\end{align}}\noindent
	where $Z_{w}\triangleq h_{jw}^2$. Thus, $\mathbb{P}_{AF}$ can be expressed as 
	{\small \begin{align}
	&\mathbb{P}_{AF} = \Pr \left( {{\kappa ^2} + {C_{2w}}Z_W > {\varepsilon _{k}}} \right) = 1 - {F_{Z_w}}\left( {\frac{{ {\varepsilon _{k}} - {\kappa ^2}}}{{{C_{2w}}}}} \right)
	\notag\\&
	= 1 \!-\! \frac{{{m_{fw}}^{{m_{fw}} - 1}{{\left( {{\varepsilon _{k}} \!-\! {\kappa ^2}} \right)}^{{m_{fw}}}}}}{{{C_{2w}}^{{m_{fw}}}B\!\left(\! {{m_{fw}},{m_{sw}}} \!\right)\left( {{m_{sw}}\!- \!1} \right){^{{m_{fw}}}}{{\bar z}_w}^{{m_{fw}}}}}
	\notag\\& \times
	{}_2{F_1}\!\left(\! {{m_{fw}},{m_{fw}} \!+\! {m_{sw}},{m_{fw}} \!+ \!1; - \!\frac{{{m_{fw}}\left(\! {{\varepsilon _{k}} \!- \!{\kappa ^2}}\! \right)}}{{{C_{2w}}\left(\! {{m_{sw}}\! - \!1} \!\right){{\bar z}_w}}}} \!\right).
	\end{align}}\noindent
	On the other hand, for the term of $\mathbb{P}_{MD}$, we have
	{\small \begin{equation}
	\mathbb{P}_{MD}\! = \!\Pr \left( {{\kappa ^2} \!+\! {C_{1w}}X_w\! +\! {C_{2w}}Z_w \!<\! {\varepsilon _{k}}} \right)\! = \Pr \left( {T_w \!+ \!{C_{1w}}X_w \!<\! {\varepsilon _{k}}} \right).
	\end{equation}}\noindent
	{\color{black} Let $Y_w=C_{1w}X_w+T_w$. The CDF of $Y_w$ can be expressed as}
	{\small \begin{equation}\label{CDFYW}
	{F_{Y_w}}\left( y \right) = \int_0^\infty  {{F_{T_w}}\left( {y - t} \right)\frac{1}{{{C_{1w}}}}{f_{X_w}}\left( {\frac{t}{{{C_{1w}}}}} \right){\rm d}t}.
	\end{equation}}\noindent
	Substituting \eqref{CDFTW} and \eqref{PDFFTR} into \eqref{CDFYW}, with the help of \cite[eq. (9.113)]{gradshteyn2007}, \cite[eq. (8.331.1)]{gradshteyn2007} and \cite[eq. (01.03.07.0001.01)]{web}, we can re-write the CDF of $Y_w$ as \eqref{salfklkaeo}, shown at the bottom of the next page. 
	%----------------------------------------------------------------------------------------------------------------------------
	%----------------------------------------------------------------------------------------------------------------------------
	\newcounter{mycount7}
	\begin{figure*}[b]
	\normalsize
	\setcounter{mycount7}{\value{equation}}
	\hrulefill
	\vspace*{4pt}
	{\small \begin{align}\label{salfklkaeo}
	&{F_{Y_w}}\!\left( y \right)=\! \frac{{{\Gamma ^{ - 1}}\left( {{m_{fw}} + {m_{sw}}} \right){m_{fw}}^{{m_{fw}}}}}{{{C_{2w}}^{{m_{fw}}}B\left( {{m_{fw}},{m_{sw}}} \right){{\left( {{m_{sw}} - 1} \right)}^{{m_{fw}}}}{{\bar z}_w}^{{m_{fw}}}}}\frac{{{m_w}^{{m_w}}}}{{\Gamma \left( {{m_w}} \right)}}\sum\limits_{j = 0}^\infty  {\frac{{{K_w}^j{\alpha _{wj}}}}{{j!\Gamma \left( {j + 1} \right){{\left( {2{C_{1w}}\sigma _w^2} \right)}^{j + 1}}}}} {\left( {\frac{1}{{2\pi i}}} \right)^2}
	\notag\\&\times\!\!
	\int_{{{\cal L}_1}} {\int_{{{\cal L}_2}} {\frac{{\Gamma \left( {{m_{fw}} + {s_1}} \right)\Gamma \left( {{m_{fw}} + {m_{sw}} + {s_1}} \right)\Gamma \left( { - {s_1}} \right)}}{{\Gamma \left( {{m_{fw}} + 1 + {s_1}} \right){\Gamma ^{ - 1}}\left( { - {s_2}} \right)}}} {{\left( {\frac{{{m_{fw}}}}{{{\Omega _w}}}} \right)}^{{s_1}}}} {\left( {\frac{1}{{2{C_{1w}}\sigma _w^2}}} \right)^{{s_2}}}\int_0^\infty  {{t^{j + {s_2}}}{{\left( {y - t - {\kappa ^2}} \right)}^{{s_1} + {m_{fw}}}}{\rm d}t{\rm d}{s_2}} {\rm d}{s_1}
	\end{align}}\noindent
	\setcounter{equation}{\value{mycount7}}
	\end{figure*}
	\addtocounter{equation}{1}
	%----------------------------------------------------------------------------------------------------------------------------
	%----------------------------------------------------------------------------------------------------------------------------
	
	With the help of \cite[eq. (3.194.1)]{gradshteyn2007} and \cite[eq. (8.384.1)]{gradshteyn2007}, let ${s_1} \to  - {t_1} - {m_{fw}}$ and ${s_2} \to  - {t_2} - j - 1$. With the help of \cite[eq. (A-1)]{mathai2009h}, we have
	{\small {\color{black} \begin{align}
	&{F_{Y_w}}\left( y \right)=	\frac{{{m_w}^{{m_w}}}}{{\Gamma \left( {{m_{fw}}} \right)\Gamma \left( {{m_{sw}}} \right)\Gamma \left( {{m_w}} \right)}}\sum\limits_{j = 0}^M  {\frac{{{K_w}^j{\alpha _{wj}}}}{{j!}{\Gamma \left( {j \!+\! 1} \right)}}}\notag\\ &\times\!\!
	H_{1,0:1,3:1,1}^{0,0:3,0:1,1}\!\!\left(\!\!\!\!\! {\left. {\begin{array}{*{20}{c}}
	{\frac{{{\Omega _w}}}{{{m_{fw}}\left( {{\varepsilon _{th}}\!-\! {\kappa ^2}} \right)}}}\\
	{\frac{{2{C_{1w}}\sigma _w^2}}{{y - {\kappa ^2}}}}
	\end{array}} \!\!\!\!\right|\!\!\!\!\begin{array}{*{20}{c}}
	{\left( {1;\!1,\!1} \right)\!\!:\!\!\left( {1\!- \!{m_{fw}},1} \right)\!\left( {1,\!1} \right)\!\!:\!\!\left( { - \!j,\!1} \right)}\\
	{ - \!:\!\left( {0,1} \right)\!\left( {{m_{sw}},1} \right)\left( {1,1} \right)\!:\!\left( {0,1} \right)}
	\end{array}} \!\!\!\!\right)\!.
	\end{align}}}\noindent
	Therefore, we can derive $\mathbb{P}_{MD}$ as ${\mathbb{P}_{MD}} = \Pr \left( {T_w + {C_{1w}}X_w < {\varepsilon _{k}}} \right) = {F_Y}\left( {{\varepsilon _{k}}} \right)$ to complete the proof.
	
	\section{Proof of Theorem \ref{IEEEPROOF2}}\label{appendix2} 
	\renewcommand{\theequation}{D-\arabic{equation}}
	\setcounter{equation}{0}
	First, the covert rate can be derived as
	{\small \begin{equation}\label{aefjeal}
	R_k = \int_0^\infty  {\log_2 \left( {1 + \gamma } \right){f_{\gamma_k} }\left( \gamma  \right){\rm d}\gamma}.
	\end{equation}}\noindent
	Substituting \eqref{EQ-PDF} into \eqref{aefjeal}, we obtain the expression of $R_k$ as \eqref{sajofeo}, shown at the bottom of the next page, where $I_{C}=\int_0^\infty  {\log_2 \left( {1 + \gamma } \right){\gamma ^{{t_1} - 1}}{\rm d}\gamma }$.
	%----------------------------------------------------------------------------------------------------------------------------
	%----------------------------------------------------------------------------------------------------------------------------
	\newcounter{mycount41}
	\begin{figure*}[b]
	\normalsize
	\setcounter{mycount41}{\value{equation}}
	\hrulefill
	\vspace*{4pt}
	{\small {\color{black} \begin{align}\label{sajofeo}
	R_k =& \frac{{{m_k}^{{m_k}}}}{{\Gamma \left( {{m_k}} \right)\Gamma \left( {{m_{fk}}} \right)\Gamma \left( {{m_{sk}}} \right)}}{\left( {\frac{{{m_{fk}}{\kappa ^2}}}{{{\Omega _k}}}} \right)^{{m_{fk}}}}\sum\limits_{j = 0}^\infty  {\frac{{{K_k}^j{\alpha _{kj}}}}{{\Gamma \left( {j + 1} \right)j!}}} {\left( {\frac{1}{{2\pi i}}} \right)^2}\int_{{{\cal L}_1}} {\int_{{{\cal L}_2}} {{{\left( {\frac{{{\Omega _k}}}{{{m_{fk}}{\kappa ^2} - {\Omega _k}}}} \right)}^{{s_2}}}} } 
	\notag\\&\times\!
	\frac{{\Gamma \left( {1 + j - {s_1}} \right)\Gamma \left( {{m_{sk}} - {s_1}} \right)\Gamma \left( {{m_{fk}} - {s_2}} \right)\Gamma \left( {{m_{sk}} + {m_{fk}} - {s_2}} \right)\Gamma \left( {{s_2}} \right)}}{{\Gamma \left( {{m_{fk}} + {m_{sk}} - {s_1} - {s_2}} \right)}}{\left( {\frac{{{\kappa ^2}}}{{2\sigma _k^2}}\frac{1}{{{C_{1k}}}}} \right)^{{s_1}}}{I_C}{\rm d}{s_2}{\rm d}{s_1},
	\end{align}}}\noindent
	\setcounter{equation}{\value{mycount41}}
	\end{figure*}
	\addtocounter{equation}{1}
	%----------------------------------------------------------------------------------------------------------------------------
	%----------------------------------------------------------------------------------------------------------------------------
	
	With the help of \cite[eq. (2.6.9.21)]{Prudnikov1986Integrals} and \cite[eq. (8.334.3)]{gradshteyn2007}, $I_{C}$ can be deduced as
	{\small \begin{equation}\label{iczhel}
	I_{C}=\int_0^\infty  {\log_2 \left( {1 + \gamma } \right){\gamma ^{{s_1} - 1}}{\rm d}\gamma }  = \frac{{ - 1}}{{\ln 2}}\Gamma \left( {{s_1}} \right)\Gamma \left( { - {s_1}} \right).
	\end{equation}}\noindent
	Substituting \eqref{iczhel} into \eqref{sajofeo} and using \cite[eq. (A-1)]{mathai2009h}, we can obtain \eqref{EQ-1}, which completes the proof.
	
	{\color{black} \section{Proof of Lemma \ref{appendixDD}}\label{appendixD} 
		\renewcommand{\theequation}{E-\arabic{equation}}
		\setcounter{equation}{0}
		Using \cite[eq. (01.03.07.0001.01)]{web} and exchanging the order of integration, we can express $I_L$ as $ {I_L} = \frac{1}{{2\pi i}}\int_{\cal L} {\Gamma \left( { - s} \right)} {D^s}{I_D}ds $, where $ {I_D} = \int_{{T_1}}^{{T_2}} {{t^{A + s}}{{\left( {B - t} \right)}^C}} dt $.  
		For different cases of $T_1$ and $T_2$, $I_D$ can be solved with the help of \cite[eq. (3.194.1)]{gradshteyn2007}, \cite[eq. (06.19.07.0001.01)]{web}, \cite[eq. (3.194.2)]{gradshteyn2007}, and \cite[eq. (3.191.1)]{gradshteyn2007} as
		{\small \begin{equation}
				{\begin{array}{*{20}{l}}
						{{\rm{Cases}}}&{{\rm{Results}}}\\\!\!\!
						\begin{array}{l}
							{T_1} = 0,\\
							{T_2} = T
						\end{array}&\!\!{{I_{{D_1}}}\!\! =\! \frac{{{B^C}{T_2}^{1 + A + s}}}{{1 + A + s}}{}_2{F_1}\!\left( { - C,1 \!+\!A \!+ \!s;2 \!+\! A \!+\! s;\frac{{{T_2}}}{B}} \right)},\\\!\!\!
						\begin{array}{l}
							{T_1} = T,\\
							{T_2} = \infty 
						\end{array}&\!\!{{I_{{D_2}}}\!\! =\! \frac{{{B^{1 + A + C + s}}}}{{{{\left( { - 1} \right)}^C}}}B\!\left(\! {\frac{B}{{{T_1}}}, -\! 1\! -\! A \!-\! C \!-\! s,1 \!+ \!C} \!\right)},\\\!\!\!
						\begin{array}{l}
							{T_1} = 0,\\
							{T_2} = \infty 
						\end{array}&\!\!{{I_{{D_3}}}\!\! =\! \frac{{\Gamma \left( { - 1 - A - C - s} \right)}}{{\Gamma \left( { - C} \right){\Gamma ^{ - 1}}\left( {1 + A + s} \right)}}{{\left( { - B} \right)}^{1 + s + A}}{B^C}},\\\!\!\!
						\begin{array}{l}
							{T_1} = 0,\\
							{T_2} = B
						\end{array}&\!\!{{I_{{D_4}}}\!\! =\! \frac{{\Gamma \left( {1 + C} \right)\Gamma \left( {1 + A + s} \right)}}{{\Gamma \left( {2 + A + C + s} \right)}}{B^{1 + s + A + C}}}.
				\end{array}}
		\end{equation}}\noindent
		Substituting $I_D$ into $I_L$, and using \cite[eq. (9.113)]{gradshteyn2007}, \cite[eq. (06.19.07.0005.01)]{web}, \cite[eq. (07.33.07.0002.01)]{web} and \cite[eq. (A-1)]{mathai2009h}, we can obtain \eqref{intr}.
		
		For the third case, i.e., $T_1=0$ and $T_2=\infty$, we have $ {I_{{L_3}}} = \Gamma \left( {1 + A} \right){B^C}{\left( { - B} \right)^{1 + A}}U\left( {1 + A,2 + A + C, - BD} \right) $, where the tricomi confluent hypergeometric function $U$ can be approximated by \cite[eq. (07.33.06.0012.01)]{web} and \cite[eq. (07.33.06.0004.01)]{web}
		{\small \begin{equation}
				\left\{\!\! \begin{array}{l}\!
					U\!\left( {a,b,c} \right) = {c^{ - a}}\left( {1 + \mathcal{O}\left( {\frac{1}{c}} \right)} \right), c \to \infty, \noindent\\
					\! U\!\left( {a,b,c} \right) \!= \!\left(\!\! {\frac{{\Gamma\!\left( {1 - b} \right)}}{{\Gamma\!\left( {a - b + 1} \right)}} \!+\! \frac{{\Gamma\!\left( {b - 1} \right)}}{{\Gamma\!\left( a \right)}}{c^{1 - b}}} \!\right)\!\left( {1 \!+\! \mathcal{O}\left( c \right)} \right), c \to 0,
				\end{array} \right.
		\end{equation}}\noindent
		which completes the proof.}

	\section{Proof of Theorem \ref{appendixEE}}\label{appendixE} 
	\renewcommand{\theequation}{F-\arabic{equation}}
	\setcounter{equation}{0}
	Obviously, the value of  ${\varepsilon _{k}}$ should be large, at least larger than ${\kappa ^2}$. The reason is that the false alarm probability is always equal to $1$ if ${\varepsilon _{k}} < {\kappa ^2}$.
	With the help of \cite[eq. (9.100)]{gradshteyn2007} and \eqref{salfklkaeo}, taking the second order derivative of ${\xi _k}$, we obtain \eqref{detection21}, shown at the bottom of the next page.
	%----------------------------------------------------------------------------------------------------------------------------
	%----------------------------------------------------------------------------------------------------------------------------
	\newcounter{mycount431}
	\begin{figure*}[b]
	\normalsize
	\setcounter{mycount431}{\value{equation}}
	\hrulefill
	\vspace*{4pt}
	{\small {\color{black} \begin{align}\label{detection21}
	\frac{{{\partial ^2}{\xi _k}}}{{\partial {\varepsilon _{th}}^2}} =& \frac{{{m_{fw}}^{{m_{fw}} - 1}}}{{{C_4}^{{m_{fw}}}B\left( {{m_{fw}},{m_{sw}}} \right)\left( {{m_{sw}} - 1} \right){{\bar z}_w}^{{m_{fw}}}}}\frac{{{m_w}^{{m_w}}}}{{\Gamma \left( {{m_w}} \right)}}\sum\limits_{j = 0}^\infty  {\frac{{{K_w}^j{\alpha _{wj}}}}{{j!}}} \frac{{\left( {q + {m_{fw}}} \right)\left( {q + {m_{fw}} - 1} \right)}}{{\Gamma \left( {j + 1} \right){{\left( {2{C_3}\sigma _w^2} \right)}^{j + 1}}}}
	\notag\\&\times
	\int_0^\infty  {{t^j}{{\left( {{\varepsilon _{th}} - {\kappa ^2} - t} \right)}^{{m_{fw}}}}\exp \left( { - \frac{t}{{2{C_3}\sigma _w^2}}} \right)\sum\limits_{q = 0}^\infty  {\frac{{{{\left( {{m_{fw}}} \right)}_q}{{\left( {{m_{fw}} + {m_{sw}}} \right)}_q}}}{{{{\left( {{m_{fw}} + 1} \right)}_q}q!}}} {{\left( { - \frac{{{m_{fw}}\left( {{\varepsilon _{th}} - t - {\kappa ^2}} \right)}}{{{C_4}\left( {{m_{sw}} - 1} \right){{\bar z}_w}}}} \right)}^q}} dt	
	\notag\\&
	- \frac{{\left( {q + {m_{fw}}} \right)\left( {q + {m_{fw}} - 1} \right){{\left( {{\varepsilon _{th}} - {\kappa ^2}} \right)}^{{m_{fw}} - 2}}{m_{fw}}^{{m_{fw}} - 1}}}{{{C_4}^{{m_{fw}}}B\left( {{m_{fw}},{m_{sw}}} \right)\left( {{m_{sw}} - 1} \right){{\bar z}_w}^{{m_{fw}}}}}\sum\limits_{q = 0}^\infty  {\frac{{{{\left( {{m_{fw}}} \right)}_q}{{\left( {{m_{fw}} + {m_{sw}}} \right)}_q}}}{{{{\left( {{m_{fw}} + 1} \right)}_q}q!}}} {\left( { - \frac{{{m_{fw}}\left( {{\varepsilon _{th}} - {\kappa ^2}} \right)}}{{{C_4}\left( {{m_{sw}} - 1} \right){{\bar z}_w}}}} \right)^q}
	\end{align}}}\noindent
	\setcounter{equation}{\value{mycount431}}
	\end{figure*}
	\addtocounter{equation}{1}
	%----------------------------------------------------------------------------------------------------------------------------
	%----------------------------------------------------------------------------------------------------------------------------
	
	In the following, we aim to prove that $\frac{{{\partial ^2}{\xi _k}}}{{\partial {\varepsilon _{k}}^2}}>0$ for any $q$. After some algebraic manipulations, we only need to prove
	{\small \begin{equation}
	\frac{{{m_w}^{{m_w}}}}{{\Gamma \left( {{m_w}} \right)}}\sum\limits_{j = 0}^\infty  {\frac{\Gamma^{-1}\!\left( {j\! +\! 1} \right){{K_w}^j{\alpha _{wj}}{I_E}}}{{j!{{\left( {2{C_{1w}}\sigma _w^2} \right)}^{j + 1}}}}}  > {\left( {{\varepsilon _{th}} \!- \!{\kappa ^2}} \right)^{q +\! {m_{fw}} - 2}},
	\end{equation}}\noindent
	where ${I_E}=\int_0^\infty  {{t^j}{{\left( {{\varepsilon _{k}} - {\kappa ^2} - t} \right)}^{q+{m_{fw}} - 2}}\exp \left( { - \frac{t}{{2{C_{1w}}\sigma _w^2}}} \right)} {\rm d}t$. Let $T_E$ denote a constant, where ${T_E} \in \left( {0,{\varepsilon _k} - {\kappa ^2}} \right)$. Because ${{t^j}{{\left( {{\varepsilon _{k}} - {\kappa ^2} - t} \right)}^{q+{m_{fw}} - 2}}\exp \left( { - \frac{t}{{2{C_{1w}}\sigma _w^2}}} \right)}>0$ when $t< {{\varepsilon _{k}} - {\kappa ^2} }$, we have
	{\small \begin{equation}
	{I_1}>\int_{T_E}^\infty  {{t^j}{{\left( {{\varepsilon _{k}} - {\kappa ^2} - t} \right)}^{q+{m_{fw}} - 2}}\exp \left( { - \frac{t}{{2{C_{1w}}\sigma _w^2}}} \right)} {\rm d}t.
	\end{equation}}\noindent
	With the help of Lemma \ref{appendixDD}, when $T_E$ is close to zero, we only need to prove that
	{\small \begin{equation}\label{vaead}
	\frac{{{m_w}^{{m_w}}}}{{\Gamma\!\left( {{m_w}} \right)}}\!\sum\limits_{j = 0}^M \!{\frac{{{K_w}^j{\alpha _{wj}}}}{{j!}}} {\left( {{\varepsilon _k}\! -\! {\kappa ^2}} \right)^{q +\! {m_{fw}} \!- 2}}\!\geqslant\! {\left( {{\varepsilon _{th}}\! -\! {\kappa ^2}} \right)^{q + {m_{fw}} - 2}}\!.
	\end{equation}}\noindent
	Obviously, \eqref{vaead} is correct because $\frac{{{m_w}^{{m_w}}}}{{\Gamma \left( {{m_w}} \right)}}\sum\limits_{j = 0}^M {\frac{{{K_w}^j{\alpha _{wj}}}}{{j!}}}=1$. Thus, we conclude that the detection error probability is convex with respect to ${\varepsilon _{k}}$.

	\section{Proof of Theorem \ref{appendixFF}}\label{appendixF} 
	\renewcommand{\theequation}{G-\arabic{equation}}
	\setcounter{equation}{0}
	Following the similar methods in Appendix \ref{appendixE}, we can express the second order derivative of $ {{\xi _k}} $ as \eqref{dao2w}, shown at the bottom of the next page.
	%----------------------------------------------------------------------------------------------------------------------------
	%----------------------------------------------------------------------------------------------------------------------------
	\newcounter{mycount4341}
	\begin{figure*}[b]
	\normalsize
	\setcounter{mycount4341}{\value{equation}}
	\hrulefill
	\vspace*{4pt}
	{\small {\color{black} \begin{align}\label{dao2w}
	\frac{{{\partial ^2}{\xi _k}}}{{\partial {C_{2w}}^2}} =	& \frac{{\left( {q + {m_{fw}}} \right)\left( {1 + q + {m_{fw}}} \right){m_{fw}}^{{m_{fw}} - 1}}}{{{C_{2w}}^{q + {m_{fw}} + 2}B\left( {{m_{fw}},{m_{sw}}} \right)\left( {{m_{sw}} - 1} \right){^{{m_{fw}}}}{{\bar z}_w}^{{m_{fw}}}}}\frac{{{m_w}^{{m_w}}}}{{\Gamma \left( {{m_w}} \right)}}\sum\limits_{j = 0}^M {\frac{{{K_w}^j{\alpha _{wj}}}}{{j!}}} \frac{1}{{\Gamma \left( {j + 1} \right){{\left( {2{C_{1w}}\sigma _w^2} \right)}^{j + 1}}}}
	\notag\\&\times
	\int_0^\infty  {{t^j}{{\left( {{\varepsilon^{opt}_k} - {\kappa ^2} - t} \right)}^{{m_{fw}}}}\exp \left( {\frac{{ - t}}{{2{C_{1w}}\sigma _w^2}}} \right)\sum\limits_{q = 0}^\infty  {\frac{{{{\left( {{m_{fw}}} \right)}_q}{{\left( {{m_{fw}} + {m_{sw}}} \right)}_q}}}{{{{\left( {{m_{fw}} + 1} \right)}_q}q!}}} {{\left( { - \frac{{{m_{fw}}\left( {{\varepsilon^{opt}_k} - t - {\kappa ^2}} \right)}}{{\left( {{m_{sw}} - 1} \right){{\bar z}_w}}}} \right)}^q}} {\rm d}t
	\notag\\&
	- \!\frac{{\left( {q\! + \!{m_{fw}}} \right)\left(\! {1 \!+\! q\! +\! {m_{fw}}} \!\right){m_{fw}}^{{m_{fw}} - 1}{{\left( {{\varepsilon^{opt}_k} \!-\! {\kappa ^2}} \right)}^{{m_{fw}}}}}}{{{C_{2w}}^{q \!+ \!{m_{fw}} \!+\! 2}B\!\left( {{m_{fw}},{m_{sw}}} \right)\left( {{m_{sw}}\! -\! 1} \right){^{{m_{fw}}}}{{\bar z}_w}^{{m_{fw}}}}}\sum\limits_{q = 0}^\infty  {\frac{{{{\left( {{m_{fw}}} \right)}_q}{{\left( {{m_{fw}} \!+\! {m_{sw}}} \right)}_q}}}{{{{\left( {{m_{fw}}\! +\! 1} \right)}_q}q!}}} {\left( { -\! \frac{{{m_{fw}}\left( {{\varepsilon^{opt}_k} \!-\! {\kappa ^2}} \right)}}{{\left( {{m_{sw}} \!-\! 1} \right){{\bar z}_w}}}} \right)^q}
	\end{align}}}\noindent
	\setcounter{equation}{\value{mycount4341}}
	\end{figure*}
	\addtocounter{equation}{1}
	%----------------------------------------------------------------------------------------------------------------------------
	%----------------------------------------------------------------------------------------------------------------------------

	With the help of \cite[eq. (3.383.1)]{gradshteyn2007}, \cite[eq. (8.384.1)]{gradshteyn2007}, and \cite[eq. (07.20.07.0004.01)]{web}, after some algebraic manipulations, proving $\frac{{{\partial ^2}{\xi _k}}}{{\partial {C_{2w}}^2}}<0$ is equivalent to prove that
	{\small \begin{align}
	{F_2}({C_{1w}}) &\triangleq \frac{{{m_w}^{{m_w}}}}{{\Gamma \left( {{m_w}} \right)}}\sum\limits_{j = 0}^M {\frac{{{K_w}^j{\alpha _{wj}}\Gamma \left( {{m_w} + q + 1} \right)}}{{j!\Gamma \left( {j + 1} \right)}}} \frac{1}{{2\pi i}}
	\notag\\&\times
	\int_\mathcal{L} {\frac{{\Gamma \left( {t + j + 1} \right)\Gamma \left( { - t} \right)}}{{\Gamma \left( {{m_w} + q + 1 - t} \right)}}{{\left( {\frac{{2{C_{1w}}\sigma _w^2}}{{\varepsilon _k^{opt} - {\kappa ^2}}}} \right)}^t}{\text{d}}t}  < 1,
	\end{align}}\noindent
	where the integration path of $\mathcal{L}$ goes from $\sigma_L -i\infty $ to $\sigma_L+i\infty $ and $\sigma  \in \mathbb{R}$. When $C_{1w}\to \infty$, we have
	%	{\small \begin{align}
	%	F_2(C_{1w}) \approx &\frac{{{m_w}^{{m_w}}}}{{\Gamma \left( {{m_w}} \right)}}\sum\limits_{j = 0}^M {\frac{{{K_w}^j{\alpha _{wj}}}}{{j!}}} \frac{{\Gamma \left( {{m_w} + q + 1} \right)}}{{\Gamma \left( {j + 1} \right)}}\operatorname{Res} \left\{ {\frac{{\Gamma \left( {t + j + 1} \right)\Gamma \left( { - t} \right)}}{{\Gamma \left( {{m_w} + q + 1 - t} \right)}}, - j - 1} \right\}\notag\\&
	%	= \frac{{{m_w}^{{m_w}}}}{{\Gamma \left( {{m_w}} \right)}}\sum\limits_{j = 0}^M {\frac{{{K_w}^j{\alpha _{wj}}}}{{j!}}} \frac{{\Gamma \left( {{m_w} + q + 1} \right)}}{{\Gamma \left( {{m_w} + q + 2 + j} \right)}}{\left( {\frac{{2{C_{1w}}\sigma _w^2}}{{\varepsilon _k^{opt} - {\kappa ^2}}}} \right)^{ - j - 1}} \to 0,
	%	\end{align}}\noindent
	{\small \begin{align}
	\!\!\!\!\!	{F_2}(\!{C_{1w}}\!) \!\!\approx\!\! \frac{{{m_w}^{{m_w}}}}{{\Gamma\!\left( \!{{m_w}} \!\right)}}\!\sum\limits_{j = 0}^M\!{\frac{{{K_w}^j{\alpha _{wj}}\Gamma\!\left(\! {{m_w} \!+\! q \!+\! 1} \!\right)}}{{j!\Gamma\! \left( {{m_w}\! + \!q\! + \!2\! + \!j} \right)}}} \!{\left(\!\! {\frac{{\varepsilon _k^{opt} \!-\! {\kappa ^2}}}{{2{C_{1w}}\sigma _w^2}}} \!\!\right)^{ j\!+\! 1}} \!\!\!\!\!\to\!0,
	\end{align}}\noindent
	which means that there must exist $C_{1w}'$ such that when $C_{1w}$ is greater than $C_{1w}'$, we always have $\frac{{{\partial ^2}{\xi _k}}}{{\partial {C_{2w}}^2}}<0$. This is reasonable because the UAV transmit power allocated to users should be large enough to meet the QoS requirements. Users may quit gaming if their data rate is too low. 
	
	To obtain $ {\frac{{{\partial ^2}{\xi _k}}}{{\partial {C_{1w}}\partial {C_{2w}}}}} $, $ {\frac{{{\partial ^2}{\xi _k}}}{{\partial {C_{2w}}\partial {C_{1w}}}}} $ and $ {\frac{{{\partial ^2}{\xi _k}}}{{\partial {C_{1w}}^2}}} $, with the help of Appendix E, we can rewrite ${\xi _k}$ as \eqref{aefkkok2}, shown at the bottom of the next page, where $ {I_J} = \int_0^\infty  {{t^{j + p}}{{\left( {{\varepsilon _k}-{\kappa ^2}-t} \right)}^{q + {m_{fw}}}}} {\rm d}t $.
	%----------------------------------------------------------------------------------------------------------------------------
	%----------------------------------------------------------------------------------------------------------------------------
	\newcounter{myount4341}
	\begin{figure*}[b]
	\normalsize
	\setcounter{myount4341}{\value{equation}}
	\hrulefill
	\vspace*{4pt}
	{\small {\color{black} \begin{align}\label{aefkkok2}
	& 	{\xi _k}= \!1\! - \frac{{{m_{fw}}^{{m_{fw}} - 1}{{\left( {{\varepsilon _k} - {\kappa ^2}} \right)}^{{m_{fw}}}}}}{{B\!\left( {{m_{fw}},{m_{sw}}} \right)\left( {{m_{sw}} - 1} \right){^{{m_{fw}}}}{{\bar z}_w}^{{m_{fw}}}}}\!\sum\limits_{q = 0}^\infty  \!{\frac{{{{\left( {{m_{fw}}} \right)}_q}{{\left( {{m_{fw}} \!+\! {m_{sw}}} \right)}_q}}}{{{{\left( {{m_{fw}} + 1} \right)}_q}q!}}} {\left( {  \frac{{{m_{fw}}\left( {{\varepsilon _k} - {\kappa ^2}} \right)}}{{\left( {1-{m_{sw}} } \right){{\bar z}_w}}}} \right)^q}\!\frac{1}{{C_{2w}^{q + {m_{fw}}}}} \!+\! \frac{{{m_{fw}}^{{m_{fw}} - 1}{{\bar z}_w}^{ - {m_{fw}}}}}{{B\!\left( {{m_{fw}},{m_{sw}}} \right)\!{{\left( {{m_{sw}} - 1} \right)}^{{m_{fw}}}}}}
	\notag\\&\times \frac{{{m_w}^{{m_w}}}}{{\Gamma \left( {{m_w}} \right)}}\sum\limits_{j = 0}^\infty  {\frac{{{K_w}^j{\alpha _{wj}}}}{{j!}}} \frac{1}{{\Gamma\!\left( {j\! + \!1} \right){{\left( {2\sigma _w^2} \right)}^{j + 1}}}}  \sum\limits_{p = 0}^\infty  {\frac{1}{{p!}}{{\left( {\frac{{ - 1}}{{2\sigma _w^2}}} \right)}^p}} \sum\limits_{q = 0}^\infty  {\frac{{{{\left( {{m_{fw}}} \right)}_q}{{\left( {{m_{fw}} + {m_{sw}}} \right)}_q}}}{{{{\left( {{m_{fw}} + 1} \right)}_q}q!}}} {\left( {\frac{{ - {m_{fw}}}}{{\left( {{m_{sw}} - 1} \right){{\bar z}_w}}}} \right)^q}\frac{1}{{C_{1w}^{p + j + 1}}}\frac{1}{{C_{2w}^{q + {m_{fw}}}}}{I_J}
	\end{align}}}\noindent
	\setcounter{equation}{\value{myount4341}}
	\end{figure*}
	\addtocounter{equation}{1}
	%----------------------------------------------------------------------------------------------------------------------------
	%----------------------------------------------------------------------------------------------------------------------------
	With the help of \eqref{aefkkok2}, we can obtain 
	{\small \begin{align}\label{ad29js}
	&	\frac{{{\partial ^2}{\xi _k}}}{{\partial {C_{1w}}^2}}\!=\! \frac{{{m_w}^{{m_w}}}{\Gamma^{-1}\!\left( {{m_w}} \right)}{{m_{fw}}^{{m_{fw}}\! -\! 1}}{{\bar z}_w}^{{-m_{fw}}}}{{B\!\left( {{m_{fw}},{m_{sw}}} \right)\!\left(\! {{m_{sw}} \!-\! 1}\! \right){^{{m_{fw}}}}}}{I_J}
	\notag\\&\times
	\sum\limits_{j = 0}^\infty \! {\frac{\Gamma^{-1}\!\left(\! {j \!+\! 1} \!\right){{K_w}^j{\alpha _{wj}}}}{{j!}{{{\left( {2\sigma _w^2} \right)}^{j + 1}}}}} \sum\limits_{p = 0}^\infty   \sum\limits_{q = 0}^\infty  {\frac{{{{\left( {{m_{fw}}} \right)}_q}\!{{\left( \!{{m_{fw}}\! + \!{m_{sw}}}\! \right)}_q}}}{{{{\left(\! {{m_{fw}}\! + \!1} \!\right)}_q}q!}}}
	\notag\\&\times
	{\frac{1}{{p!}}{{\left( {\frac{{ - 1}}{{2\sigma _w^2}}} \right)}^p}}{\left( {\frac{{ - {m_{fw}}}}{{\left( {{m_{sw}} - 1} \right){{\bar z}_w}}}} \right)^q}\frac{{\left( {p + j + 1} \right)\left( {p + j + 2} \right)}}{{C_{1w}^{p + j + 3}C_{2w}^{q + {m_{fw}}}}},
	\end{align}}\noindent
	and
	{\small \begin{align}\label{afaef341}
	&\frac{{{\partial ^2}{\xi _k}}}{{\partial {C_{1w}}\partial {C_{2w}}}} =\frac{{{\partial ^2}{\xi _k}}}{{\partial {C_{2w}}\partial {C_{1w}}}}= \frac{{I_J}{{\bar z}_w}^{-{m_{fw}}}{{m_{fw}}^{{m_{fw}} - 1}}}{{B\left( {{m_{fw}},{m_{sw}}} \right)\left( {{m_{sw}} - 1} \right){^{{m_{fw}}}}}}
	\notag\\&\times
	\frac{{{m_w}^{{m_w}}}}{{\Gamma\!\left( {{m_w}} \right)}}\sum\limits_{j = 0}^\infty  {\frac{{{K_w}^j{\alpha _{wj}}}}{{j!}{\Gamma\!\left( {j + 1} \right){{\left( {2\sigma _w^2} \right)}^{j + 1}}}}} \sum\limits_{p = 0}^\infty  {\frac{1}{{p!}}{{\left( {\frac{{ - 1}}{{2\sigma _w^2}}} \right)}^p}}
	\notag\\&\times\!
	\sum\limits_{q = 0}^\infty \! {\frac{{{{\left(\! {{m_{fw}}} \!\right)}_q}{{\left(\! {{m_{fw}} \!+\! {m_{sw}}} \!\right)}_q}}}{{{{\left(\! {{m_{fw}}\! + \!1} \!\right)}_q}q!}}} \!{\left(\!\! {\frac{{ {{\bar z}_{w}^{-1}} {m_{fw}}}}{{{ 1\!-\!{m_{sw}} }}}} \!\!\right)^q}\!\frac{{p\! + \!j \!+\! 1}}{{C_{1w}^{p \!+\! j\! + \!2}}}\frac{{q\! + \!{m_{fw}}}}{{C_{2w}^{q \!+ \!{m_{fw}} \!+ \!1}}}.
	\end{align}}\noindent
	Comparing \eqref{ad29js} and \eqref{afaef341} with \eqref{aefkkok2}, we can conclude that$ \frac{{{\partial ^2}{\xi _k}}}{{\partial {C_{1w}}\partial {C_{2w}}}} \times \frac{{{\partial ^2}{\xi _k}}}{{\partial {C_{2w}}\partial {C_{1w}}}}>0$ and $\frac{{{\partial ^2}{\xi _k}}}{{\partial {C_{1w}}^2}}>0$. Thus, we have $ {\nabla ^2}{\xi _k} = \frac{{{\partial ^2}{\xi _k}}}{{\partial {C_{1w}}^2}}\frac{{{\partial ^2}{\xi _k}}}{{\partial {C_{2w}}^2}} - \frac{{{\partial ^2}{\xi _k}}}{{\partial {C_{1w}}\partial {C_{2w}}}}\frac{{{\partial ^2}{\xi _k}}}{{\partial {C_{2w}}\partial {C_{1w}}}} < 0 $, which completes the proof.
	
	\end{appendices}
	\bibliographystyle{IEEEtran}
	\bibliography{IEEEabrv,Ref}
	%\end{thebibliography}
	
	\end{document}